\documentclass[reprint, aps, prd, superscriptaddress, longbibliography, nofootinbib]{revtex4-2}

\usepackage{amsmath}
\usepackage{amssymb}
\usepackage{graphicx}
\usepackage{tabularx}
\usepackage{siunitx}
\usepackage{hyperref}

\newcommand{\ra}{^{\mathrm{h}}}
\newcommand{\dec}{^{\circ}}
\newcommand{\cA}{\mathcal{A}}
\newcommand{\cF}{\mathcal{F}}
\newcommand{\hcF}{\hat{\cF}}
\newcommand{\fmin}{f_{\mathrm{min}}}
\newcommand{\fmax}{f_{\mathrm{max}}}
\newcommand{\fd}{\dot{f}}
\newcommand{\fdmin}{\fd_{\mathrm{min}}}
\newcommand{\fdmax}{\fd_{\mathrm{max}}}
\newcommand{\mcoh}{\tilde{\mu}_{\mathrm{max}}}
\newcommand{\minc}{\hat{\mu}_{\mathrm{max}}}
\newcommand{\SH}{S_{\mathrm{H}}}
\newcommand{\SL}{S_{\mathrm{L}}}
\newcommand{\Izz}{I_{\mathrm{zz}}}

\begin{document}

\title{Deep exploration for continuous gravitational waves at 171--172 Hz in LIGO second observing run data}

\author{Karl Wette}
\email{karl.wette@anu.edu.au}
\affiliation{Centre for Gravitational Astrophysics, Australian National University, Canberra ACT 2601, Australia}
\affiliation{ARC Centre of Excellence for Gravitational Wave Discovery (OzGrav), Hawthorn VIC 3122, Australia}

\author{Liam Dunn}
\affiliation{School of Physics, University of Melbourne, Parkville VIC 3010, Australia}
\affiliation{ARC Centre of Excellence for Gravitational Wave Discovery (OzGrav), Hawthorn VIC 3122, Australia}

\author{Patrick Clearwater}
\affiliation{Gravitational Wave Data Centre, Swinburne University of Technology, Hawthorn VIC 3122, Australia}
\affiliation{School of Physics, University of Melbourne, Parkville VIC 3010, Australia}
\affiliation{ARC Centre of Excellence for Gravitational Wave Discovery (OzGrav), Hawthorn VIC 3122, Australia}

\author{Andrew Melatos}
\affiliation{School of Physics, University of Melbourne, Parkville VIC 3010, Australia}
\affiliation{ARC Centre of Excellence for Gravitational Wave Discovery (OzGrav), Hawthorn VIC 3122, Australia}

\date{\today}

\begin{abstract}
We pursue a novel strategy towards a first detection of continuous gravitational waves from rapidly-rotating deformed neutron stars.
Computational power is focused on a narrow region of signal parameter space selected by a strategically-chosen benchmark.
We search data from the 2nd observing run of the LIGO Observatory with an optimised analysis run on graphics processing units.
While no continuous waves are detected, the search achieves a sensitivity to gravitational wave strain of $h_0 = \num{1.01e-25}$ at 90\% confidence, 24\% to 69\% better than past searches of the same parameter space.
Constraints on neutron star deformity are within theoretical maxima, thus a detection by this search was not inconceivable.
\end{abstract}

\maketitle

\section{Introduction}

Neutron stars, the dense remnants of exploded stars, are of particular interest in gravitational wave astronomy.
Two orbiting neutron stars inevitably collide and merge, generating a characteristic \emph{chirp} signal in gravitational waves detectable by the LIGO~\cite{LIGO2015:AdvLIG} and Virgo~\cite{AcerEtAl2018:SttAdvVr} observatories.
The first such detection, in gravitational waves~\cite{LIGOVirg2017:GWObsGrvWBNtSIn} and electromagnetic radiation~\cite{LIGOVirg2017:MltObsBnNtSMr}, has enlightened neutron stars physics, heavy element production in the Universe, and cosmology.

A single rapidly-rotating neutron star may also emit gravitational waves, provided it is non-axisymmetric.
The expected \emph{continuous} wave signal is long-lived, decreases in frequency over time as rotational energy is radiated in gravitational waves, and is modulated by the relative motion between neutron star and detector.
The neutron star non-axisymmetry might arise from deformations due to e.g. its magnetic field, accretion of matter from a companion star, or normal oscillation modes; see~\cite{Rile2017:RcSrCntGrvWv,SienBejg2019:CntGrvWNtSCrSPr} for recent reviews.
While theoretical and observational predictions of the non-axisymmetry exist~\cite{MelaPayn2005:GrvRdAcMlPMgnCM,MastMela2012:UGrvULIntMgFStRcP,JohnOwen2013:MxElDfrRltSt,WoanEtAl2018:EvMnEllMllPl,gittins2020modelling}, typical non-axisymmetries of Galactic neutron stars are unknown~\cite{Palo2005:SmlPpIsNSEvTEGrvW,WadeEtAl2012:CntGrvWIsGlNSAdDE}.

Continuous wave signals are expected to be marginally detectable by contemporary detectors; analysis of year-long datasets, at significant computational cost, is likely required for a first detection~\cite{BradEtAl1998:SrcPrdSrLI}.
It is uncertain what sensitivity is required, and what region of signal parameter space should be explored, in order to maximise detection prospects.
Moreover, due to finite computational resources, search sensitivity and parameter-space coverage cannot be maximised simultaneously.
This mandates the use of strategies which balance these objectives.

Depth-first strategies, which prioritise sensitivity at the expense of parameter-space coverage, are employed in searches for continuous waves from known pulsars~\cite{LIGOVirg2019:SrGrvWvKPlTHrm20152017LD,LIGOVirg2019:NrrSrGrvWKPlUSLObR,LIGOVirg2020:GrvCnsEqElMlP}.
While the signal parameters are known, the targeted pulsars may not radiate detectable continuous waves, or else their gravitational-wave and electromagnetic frequencies may differ~\cite{JoneAnde2002:GrvWvFrPrcNtS}.
Breadth-first strategies, which prioritise parameter-space coverage at the expense of sensitivity, are used in all-sky surveys for electromagnetically-quiet neutron stars~\cite{LIGOVirg2019:ASCntGrvWIsNtSUAdLOD,O2-EinsteinAtHome,CovaSint2020:FASCntGrvSgUnNSBSUAdLD}.
While the expansive parameter space may contain signals, the analysis method may be insufficiently sensitive to detect them.
Searches targeting compact remnants of supernovae~\cite{LIGOVirg2019:SrCntGrvW15SpRFmBAL,LindOwen2020:DrSCntGrvWTSpRDALSObR,MillEtAl2020:SGrvW12YSprRmHMMAdLSObR,PapaEtAl2020:SCntGrvWCCOSpRmCsVJG}, low-mass X-ray binaries~\cite{LIGOVirg2017:ULGrvWScXMdCrsSAdLD,LIGOVirg2019:SGrvWScXSAdLObsRImHMM,MiddEtAl2020:SGrvWFLMXBnSAdLObsRImHMM,zhang2020search}, and the Galactic Centre~\cite{PiccEtAl2020:DrSCntGrvSGlCALSObR}  where some parameters may be unknown, adopt intermediate strategies.

In this paper, we pursue the novel combination of an all-sky survey for continuous waves with a depth-first strategy.
The search range of gravitational-wave frequencies is limited to 1~Hz, and a single benchmark is used to select all other search parameters.
We apply an optimised analysis method to data from the 2nd observing run (O2) of LIGO\footnote{The Virgo detector joined O2 for only the last $\sim 9$\% of the run time. Due to the limited data available, which would not have noticeably improved sensitivity, Virgo O2 data was not used in this search.}~\cite{O2-data}, and utilise graphics processing units (GPUs) to maximise computational efficiency.
While no continuous wave signals are detected, our search is the most sensitive yet performed in the O2 data over the chosen parameter space, improving by 24--69\% over previous searches.

\section{Semi-coherent analysis}

The gravitational-wave strain of a continuous wave signal is written as four amplitudes $\cA_{i}$ multiplying four oscillatory basis functions~\cite{JaraEtAl1998:DAnGrvSgSpNSSDtc}.
The $\cA_{i}$ are functions of the strain amplitude $h_0$, and three angles determining the neutron star orientation and the initial signal phase.
The basis functions are functions of the phase $\phi(t)$.
For an isolated neutron star, $\phi(t)$ is a function of: the sky position, given by its right ascension $\alpha$ and declination $\delta$; and the gravitational-wave frequency $f$, and its first time derivative (or spin-down) $\fd$, as observed at the Solar System barycenter at a given reference time.
Very young ($\lesssim \SI{1000}{yr}$ old) neutron stars require higher-order frequency derivatives, and those with binary companions require additional orbital parameters.
We do not target these sources, as the consequent increase in computational burden is not justified by their expected abundances, which are comparable to isolated neutron stars.

The computational cost of continuous wave searches of large parameter spaces and year-long datasets using fully phase-coherent matched filtering would be prohibitive.
We therefore employ a semi-coherent analysis method~\cite{BradCrei2000:SrcPrSrLIIHrrSr:II,Wett2015:PrmMASmSrGrvPl,WettEtAl2018:ImpSmSCnGrvWUOCnTB}.
The data are first partitioned into $N$ segments of time-span $T$.
Within each segment, the data are filtered against a bank of signal templates.
Each template computes the $\cF$-statistic $2\cF$, the fully-coherent matched filter analytically maximised over $\cA_{i}$~\cite{JaraEtAl1998:DAnGrvSgSpNSSDtc}, which represents a signal with parameters $(\alpha', \delta', f', \fd')$.
The template bank is constructed using a parameter-space metric~\cite{WettPrix2013:FPrmMtASrGrvPl,Wett2016:EmExRVlPrmMASGrvP} to ensure any signal is recovered within a prescribed maximum loss in signal power (relative to a perfect match), known as the maximum mismatch and given by $\mcoh$.
Optimal lattices~\cite{Prix2007:TmpSrGrvWEfLCFPrS,Wett2014:LTmPlcChASrGrvP} are used to reduce the overlap between nearby templates, and thereby minimise the template bank and the cost of computing $2\cF$.

Continuous wave signal templates are then constructed spanning the whole dataset, with parameters $(\alpha, \delta, f, \fd)$ drawn from a second template bank with maximum mismatch $\minc$.
For each full-span template, and for each segment, the per-segment template is selected whose frequency evolution $f(t) = d[\phi(t)/2\pi]/dt$ most closely matches the full-span template, as determined by the parameter-space metric~\cite{Wett2015:PrmMASmSrGrvPl}.
Then, for each full-span template, we compute the detection statistic $2\hcF$: the mean of the $N$ values of $2\cF$ corresponding to the $N$ best-match per-segment templates.
This technique permits better sensitivity than a fully-coherent analysis given limited computational resources~\cite{PrixShal2012:SCntGrvWOpStMFCmC}.
The search setup is determined by the parameters $N$, $T$, $\mcoh$, and $\minc$.

\section{Parameter space}

\begin{table}
    \centering
    \begin{tabularx}{\linewidth}{l@{\extracolsep{\fill}}r}
    \hline\hline
    Parameter                            & Value \\
    \hline
    Minimum frequency $\fmin$            & \SI{171.0}{Hz} \\
    Maximum frequency $\fmax$            & \SI{172.0}{Hz} \\
    Minimum spin-down $\fdmin$           & \SI{-2.7e-13}{Hz.s^{-1}} \\
    Maximum spin-down $\fdmax$           & \SI{-3.8e-18}{Hz.s^{-1}} \\
    Number of segments $N$               & 26 \\
    Segment time-span $T$                & \SI{858194}{s} (\SI{9.9}{d}) \\
    Maximum per-segment mismatch $\mcoh$ & 0.097 \\
    Maximum full-span mismatch $\minc$   & 0.27 \\
    \hline\hline
    \end{tabularx}
    \caption{
        Parameters of the search.
    }
    \label{tab:params}
\end{table}

\begin{figure}
    \centering
    \includegraphics[width=\linewidth]{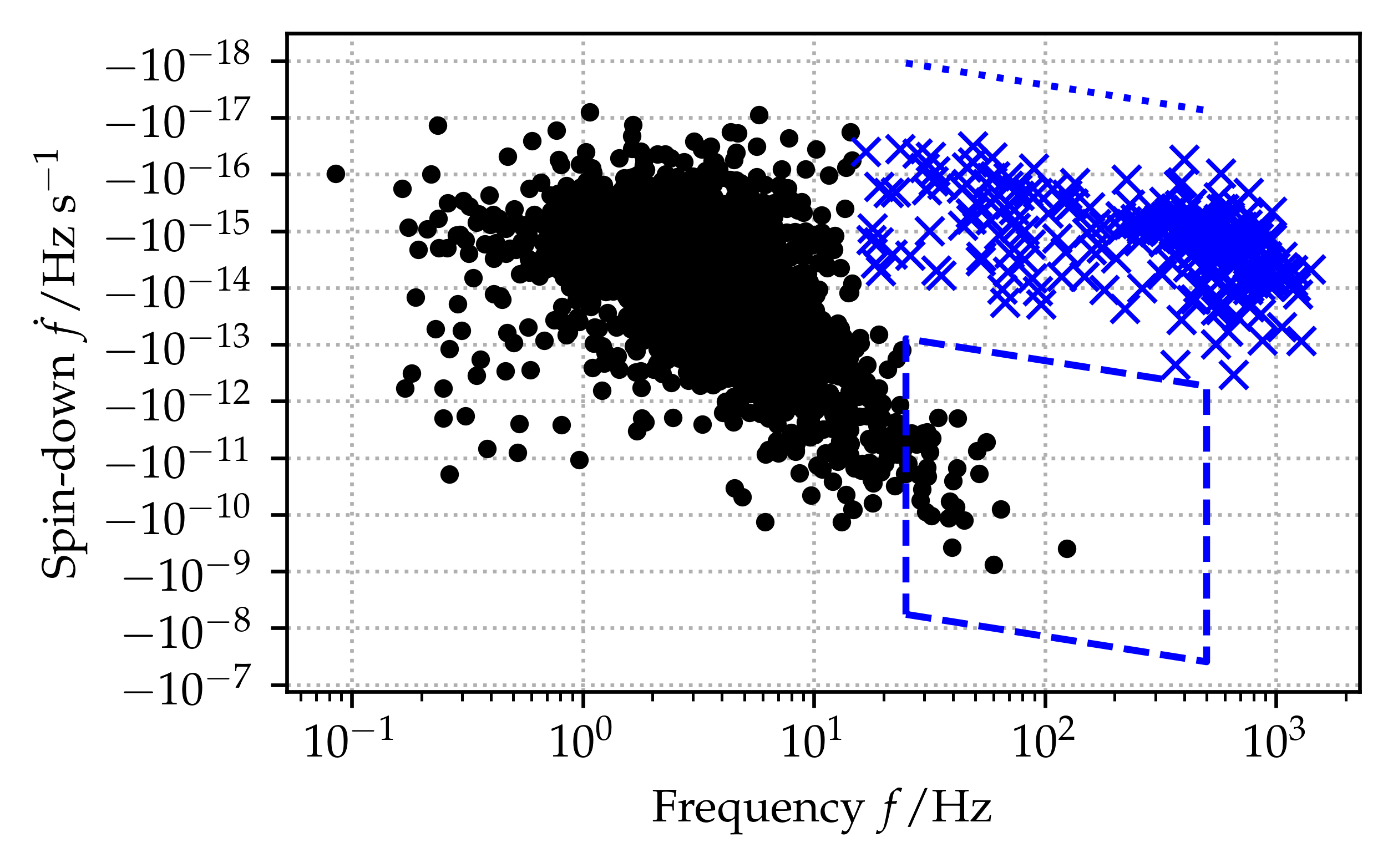}
    \caption{
        Gravitational-wave frequency $f$ versus spin-down $\fd$.
        Dots/crosses represent 2289 pulsars~\cite{MancEtAl2005:AstTlsNtFcPCt} with $\fd < 0$.
        Crosses indicate a subset of 289 pulsars with $f > \SI{15}{Hz}$ and $\fd \gtrsim \SI{-e-12}{Hz.s^{-1}}$.
        Gravitational-wave emission at twice the neutron star rotation frequency is assumed.
        The horizontal extent of the dashed/dotted lines denotes the range of $\fmin$ considered during the search setup; at a given $\fmin$, the dashed area denotes the range of $\fdmin$ considered, and the dotted line denotes $\fdmax$.
    }
    \label{fig:spindown_range}
\end{figure}

Table~\ref{tab:params} lists the parameters of the search.
They maximise the benchmark
\begin{equation}
    \label{eq:benchmark}
    h_0^{-2} \times \sigma_{h}^{-1/4} \times \rho^{1/3} \,.
\end{equation}
In the first factor, $h_0$ is the estimated sensitivity~\cite{Wett2012:EsSnWdpSrGrvP,DreiEtAl2018:FAcSnsEsCntSr} of a semi-coherent search of the LIGO O2 data with the given parameters.
The negative exponent denotes that more sensitive searches, where $h_0$ is smaller, are preferred.
In the second factor,
\begin{equation}
    \label{eq:sigma_h}
    \sigma_{h} = \frac{ \sqrt{ \sigma_{\SH}^2 + \sigma_{\SL}^2 } }{ \SH + \SL } \,,
\end{equation}
$\SH$ and $\SL$ are the noise power spectral densities, harmonically averaged over a 1-Hz band, of O2 data from the LIGO detectors at Hanford, WA, and Livingston, LA respectively; and $\sigma_{\SH}$ and $\sigma_{\SL}$ are the respective standard deviations of the 1800 power spectral density bins in the same 1-Hz band.
The exponent of $\sigma_{h}$ was chosen empirically to favour 1-Hz bands where the power spectrum has minimal variation over frequency, and does not contain any prominent instrumental artefacts~\cite{CovaEtAl2018:IdnMtNSArDSPrGrvWFTORAL}.
In the third factor, $\rho$ is the density of observed pulsars~\cite{MancEtAl2005:AstTlsNtFcPCt} within an Earth-centred sphere with radius $d$, the distance out to which the search is sensitive, given by~\cite{WettEtAl2008:SrGrvWvCssLI}
\begin{equation}
    \label{eq:dist}
    d = h_0^{-1} \sqrt{ \frac{ 5 G \Izz }{ 8 c^3 \tau} } \,,
\end{equation}
where $G$ is the gravitational constant, $\Izz = \SI{e38}{kg.m^2}$ is the principal moment of inertia of a typical neutron star, $c$ is the speed of light, and $\tau = -f / ( 4 \fdmin )$ is the characteristic spin-down timescale assuming energy loss only in gravitational waves.
As $\rho \propto d^{-3} \propto h_0^3$ in the worst case, the exponents of $h_0$ and $\rho$ in Eq.~\eqref{eq:benchmark} are chosen so that $h_0^{-2} \rho^{1/3} \propto h_0^{-1}$ and a smaller $h_0$ is preferred.

The use of $\rho$ in Eq.~\eqref{eq:benchmark} is motivated by the hypothesis that neutron stars which predominately radiate gravitational waves are found in similar regions of the Galaxy, and in overlapping regions of the $f$--$\fd$ plane (Figure~\ref{fig:spindown_range}) as observed pulsars.
We may therefore use the observed density of pulsars as a prior on the possible density of gravitational-wave emitting neutron stars.
While it may be only approximately true, this hypothesis is useful in guiding a detection strategy.
There is no evidence that gravitational-wave emitting neutron stars only occupy special regions of the Galaxy.
Simulations of Galactic neutron stars~\cite{Palo2005:SmlPpIsNSEvTEGrvW,WadeEtAl2012:CntGrvWIsGlNSAdDE} indicate that, while electromagnetic emission leads to higher spin-down rates and hence lower neutron star rotation frequencies than gravitational radiation, nevertheless the two populations overlap in the $f$--$\fd$ plane.

Assuming gravitational-wave emission at twice the neutron star rotation frequency, the number of pulsars in Figure~\ref{fig:spindown_range} with frequencies 166.5--$\SI{176.5}{Hz}$, bracketing the search frequency range (Table~\ref{tab:params}), is 0.2 per Hz.
The expected number of Galactic neutron stars ($\gtrsim \num{e8}$;~\cite{SartEtAl2010:GlcNtStISVlDstDH}) is, however, much larger than the number of observed pulsars ($\sim \num{3e3}$;~\cite{MancEtAl2005:AstTlsNtFcPCt}).
An optimistic estimate (ignoring e.g.\ selection effects) of the number of neutron stars within the search frequency range is therefore $\sim \num{7e3}$.
More pessimistic estimates, which account for the distance out to which the search is sensitive, are outlined in the Discussion.

Millisecond pulsars~\footnote{We note that, as shown in Figure~\ref{fig:spindown_range}, the setup of the search did not consider the most rapidly-spinning millisecond pulsars with $f > \SI{500}{Hz}$, due to computational restrictions.}, with $f \gtrsim \SI{100}{Hz}$, are hypothesised to have spun up by accretion of matter from a companion star; the same emission mechanism could also have built a non-axisymmetric neutron star~\cite{MelaPayn2005:GrvRdAcMlPMgnCM}.
A plausible explanation for the maximum observed millisecond pulsar spin frequency is that spin-up due to accretion is balanced by spin-down due to gravitational waves~\cite{Bild1998:GrvRdtRtAcNtS}.
If true, this suggests that at frequencies where one finds millisecond pulsars which are still undergoing accretion, one might also expect millisecond pulsars where accretion has ceased and which may be spinning down dominantly though gravitational waves.

Parameters are chosen to maximise Eq.~\eqref{eq:benchmark} through a Monte Carlo process.
Trial values are drawn according to three sampling modes.
In the first mode (chosen for 70\% of trials), parameters are sampled from large initial ranges: $\fmin$ from $25$ to $\SI{500}{Hz}$, set by the limited sensitivity of LIGO O2 data at low frequencies, and by the quadratic scaling of computational cost with frequency; $\fdmin$ from the ranges shown in Figure~\ref{fig:spindown_range} at a given $\fmin$; $N$ from 2 to 134, the maximum number of segments with $T = 2$~days in the O2 data; and $\mcoh$ and $\minc$ from 0.05 to 1.0.
(Fixed parameters are $\fmax = \fmin + \SI{1}{Hz}$, and $\fdmax$ as shown in Figure~\ref{fig:spindown_range}.)
In the second mode (chosen for 20\% of trials) parameters are sampled from ranges 90--110\% of the last accepted parameters, in order to improve upon them.
In the third mode (chosen for 10\% of trials) only $\mcoh$ and $\minc$ are sampled, from 0.05 up to the last accepted values of $\mcoh$ and $\minc$, for the purpose of absorbing any remaining computational allowance.

For sampled trial parameters to be acceptable, certain criteria must be satisfied.
The estimated sensitivity~\cite{Wett2012:EsSnWdpSrGrvP,DreiEtAl2018:FAcSnsEsCntSr} must improve upon the last accepted trial.
The estimated computational cost~\cite{WettEtAl2018:ImpSmSCnGrvWUOCnTB} must be within 10--100\% of a prescribed budget; for the second and third modes, the cost must also be greater than the last accepted trial.
At least 20,000 iterations are performed, until the estimated cost is $\ge 99$\% of the budget.
Through experimentation we found that the Monte Carlo process is robust to fine-tuning of the sampling modes.
Aside from $\fmin$, the parameters (Table~\ref{tab:params}) do not rail against their sampling limits.

\section{Implementation}

LIGO O2 data, starting at UTC 2016 November 30 17:31:57 (GPS 1164562334), are partitioned into 16 segments, followed by a gap where no usable data are present, followed by a further 10 segments, ending at UTC 2017 August 25 21:59:34 (GPS 1187733592); all segments and gaps are of time-span $T$.
The data are further divided into 12626 blocks~\cite{O2-SFT-segments} of $\SI{1800}{s}$ duration, and then Fourier transformed.

Computational efficiency was optimised using GPUs.
For this search, times to compute the $\cF$-statistic~\cite{GPU-Fstat} and the semi-coherent $2\hcF$ are reduced by factors of $\sim 240$ and $\sim 4.2$ respectively, relative to non-GPU processors.
Computation of $2\cF$ dominates the total analysis time.
The analysis ran for $\sim 5800$~days on the OzSTAR supercomputer using NVIDIA P100 type GPUs; a total of $\num{4.3e16}$ templates were analysed.

\section{Candidates}

\begin{figure}
    \centering
    \includegraphics[width=\linewidth]{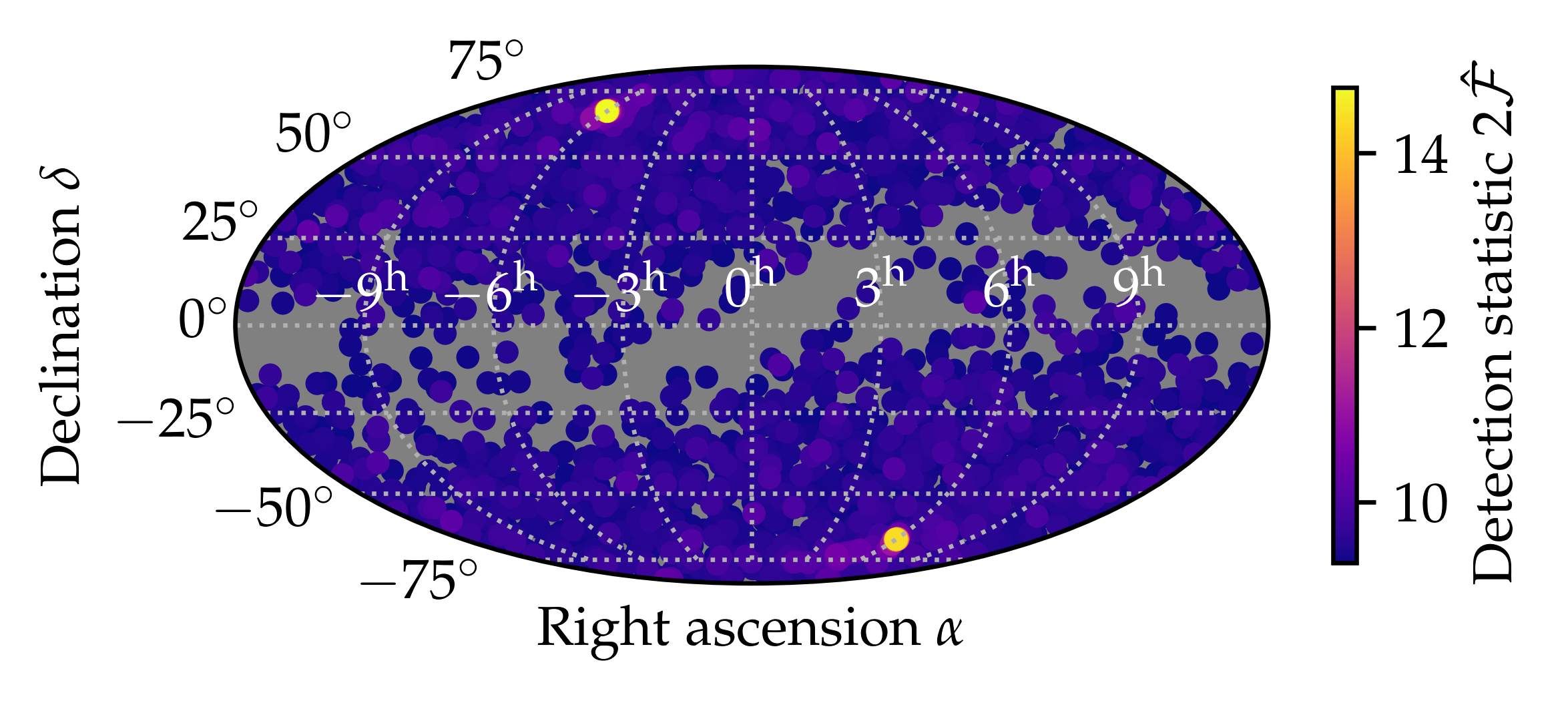}
    \caption{
        Detection statistic $2\hcF$ of the top $\num{e5}$ candidates versus right ascension $\alpha$ and declination $\delta$.
        Areas with no candidates are shaded grey.
    }
    \label{fig:results_skypos}
\end{figure}

\begin{figure}
    \centering
    \includegraphics[width=\linewidth]{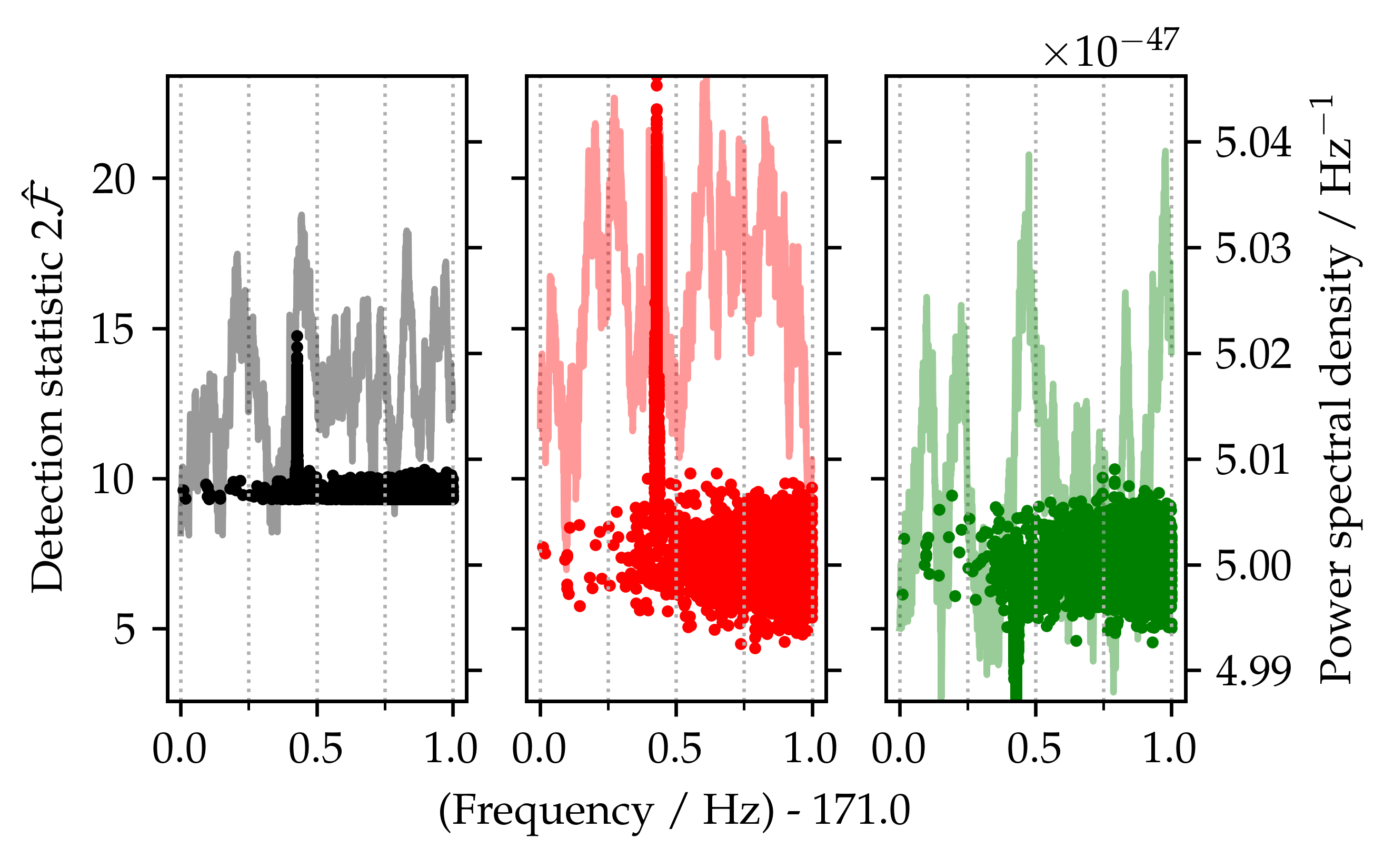}
    \caption{
        Detection statistic $2\hcF$ of the top $\num{e5}$ candidates (dark-coloured dots, left vertical axis) and the noise power spectrum of the O2 data (light-coloured lines, right vertical axis) versus frequency, for all LIGO data (left, black), and for data from the Hanford (middle, red) and Livingston (right, green) detectors.
    }
    \label{fig:results_freq}
\end{figure}

\begin{figure}
    \centering
    \includegraphics[width=\linewidth]{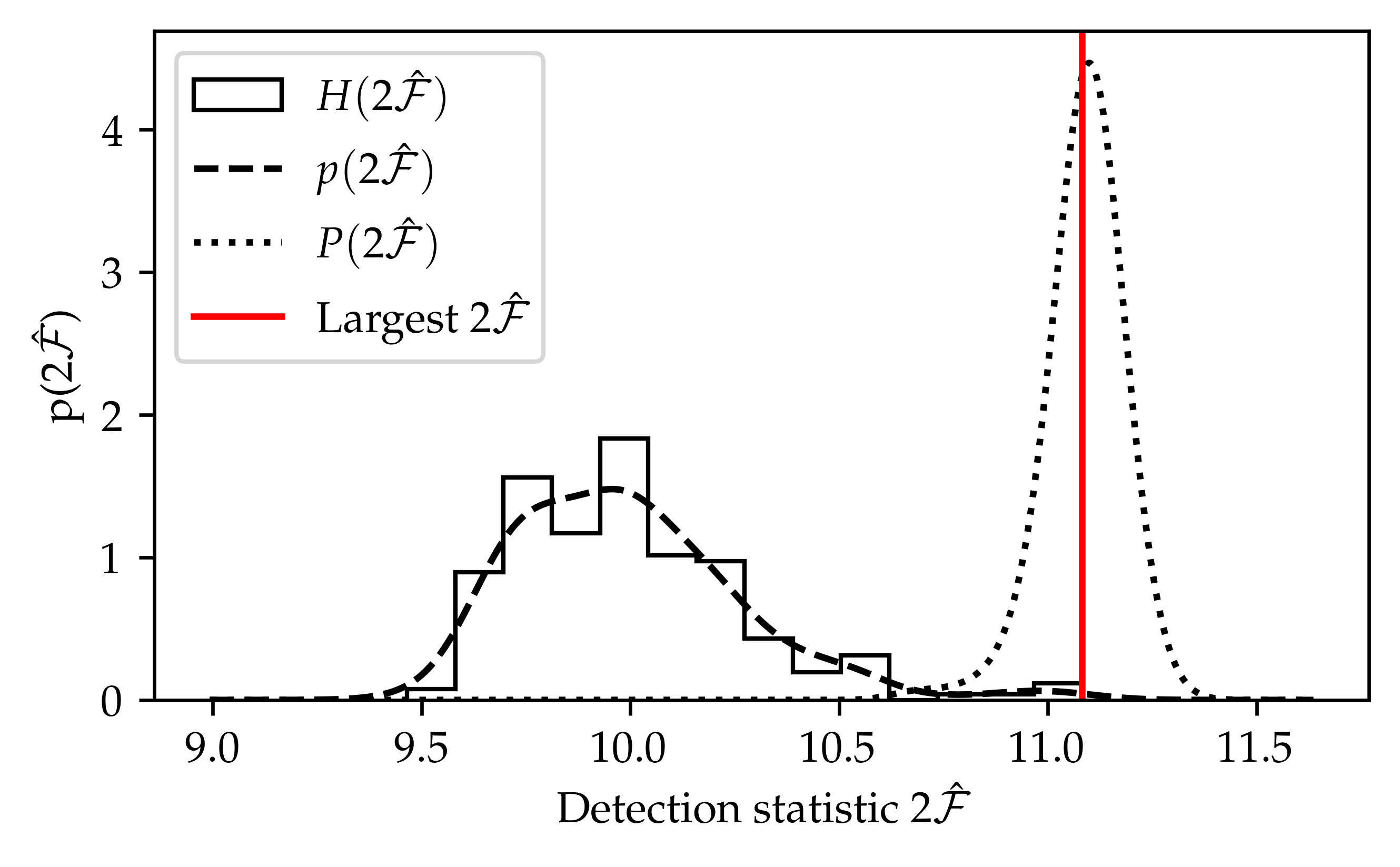}
    \caption{
        Expected distribution of the largest detection statistic $2\hcF$.
        Black solid line: histogram $H(2\hcF)$ of the largest $2\hcF$, after removal of the outlier, from 222 parameter-space partitions.
        Black dashed line: probability density $p(2\hcF)$ fit to $H(2\hcF)$.
        Black dotted line: probability density $P(2\hcF)$ of the largest $2\hcF$ from the entire search.
        Vertical red line: largest candidate $2\hcF$.
    }
    \label{fig:results_dist}
\end{figure}

\begin{table}
    \centering
    \begin{tabularx}{\linewidth}{ll@{\extracolsep{\fill}}S[table-format=1.2e-2]@{}lS[table-format=-1.1e-2]l}
    \hline\hline
    Data & Method          & {$h_0$ Sensitivity} &        & {$\fdmin$ / \si{Hz.s^{-1}}}  & Ref.                                     \\
    \hline
    O2   & $\cF$-statistic & 3.40e-25            &        & -1.0e-8                      & \cite{LIGOVirg2019:ASCntGrvWIsNtSUAdLOD} \\
    O2   & SkyHough        & 2.27e-25            &        & -1.0e-8                      & \cite{LIGOVirg2019:ASCntGrvWIsNtSUAdLOD} \\
    O1   & PowerFlux       & 2.10e-25            &        & -1.0e-8                      & \cite{DergPapa2019:SnImpSrPrGrvWUOLD}    \\
    O2   & FrequencyHough  & 1.72e-25            &        & -1.0e-8                      & \cite{LIGOVirg2019:ASCntGrvWIsNtSUAdLOD} \\
    O2   & Einstein@Home   & 1.33e-25            & $^{*}$ & -2.6e-9                      & \cite{O2-EinsteinAtHome}                 \\
    O2   & This search     & 1.07e-25            &        & -2.7e-13                                                                \\
    O2   & This search     & 1.01e-25            & $^{*}$ & -2.7e-13                                                                \\
    \hline\hline
    \end{tabularx}
    \caption{
        Estimated sensitivity $h_0$ and minimum spin-down $\fdmin$ of this search, and of previous searches of data from the 1st (O1) and 2nd (O2) LIGO observing runs which cover the same parameter space.
        All sensitivities are at 95\% confidence, except those asterisked which are at 90\% confidence.
    }
    \label{tab:h0sens}
\end{table}

Figure~\ref{fig:results_skypos} plots the detection statistic $2\hcF$ of the top $\num{e5}$ candidates as a function of sky position.
Clear outliers are visible with maximum $2\cF \approx 14.7$, at $\alpha$ separated by $\sim 12.1\ra$, and at $\delta \sim \pm 66.1\dec$.
The presence of outliers of similar strength at opposing points suggests an instrumental artefact.
Figure~\ref{fig:results_freq} plots, as a function of frequency, the single-detector $2\hcF$ of the top $\num{e5}$ candidates and the noise power spectrum of the Hanford and Livingston detectors individually, and the multi-detector $2\cF$~\cite{CutlSchu2005:GnrFsMlDtMGrvWP} and harmonically-averaged noise power spectrum of both detectors.
A feature at $f \approx \SI{171.4276}{Hz}$ appears more prominently at Hanford than at Livingston, inconsistent with an astrophysical signal but consistent with an instrumental artefact.

We perform a multi-stage follow-up search of a small region around the outlier, at negligible cost: $\alpha \in \pm [ 5.894\ra, 6.188\ra]$, $\delta \in \mp [63.517\dec, 68.599\dec]$, $f \in [171.4271, 171.4281]$~Hz.
Each follow-up search stage halves $N$, and doubles $T$, relative to the previous stage (or to the initial search), in order to improve sensitivity.
The final stage, which searches a single 160-day segment, yields a top candidate with multi-detector $2\cF = 95.1$ and single-detector $2\cF = 113.3$ and 13.33 for Hanford and Livingston respectively.
We note that the multi-detector $2\cF$ is less than one of the single-detector $2\cF$, violating a long-established consistency criterion~\cite{LIGO2009:EnsSrPrGrvWvLSD} for an astrophysical signal.
We conclude that the outlier is an instrumental artefact, and exclude candidates within the follow-up parameter space from further analysis.

Figure~\ref{fig:results_dist} plots the histogram $H(2\hcF)$ of the largest $2\hcF$, after removal of the outlier, from 222 partitions of the parameter space.
Each partition is composed of many disjoint patches evenly distributed across the sky; we expect the largest $2\hcF$ from the partitions to be statistically independent.
We fit $H(2\hcF)$ with a smooth probability density $p(2\hcF)$ using Gaussian kernel density estimation~\cite{Parz1962:EstPrbDnFnMd,Rose1956:RmSNnpEstDnFn}.
As no obvious outliers remain, we consider $p(2\hcF)$ to well-approximate the distribution of $2\cF$ in each partition in the absence of a signal.
It follows from extreme value statistics that the probability density $P(2\hcF)$ of the largest $2\hcF$ expected from the entire search, in the absence of a signal, is~\cite{LIGO2010:FrSrGrvWYngKNtS}
\begin{equation}
    \label{eq:P_2F}
    P(2\hcF) = \frac{d}{d(2\hcF)} \left[ \int_{0}^{2\hcF} d(2\hcF') p(2\hcF') \right]^{222} \,.
\end{equation}
The largest found $2\hcF = 11.1$, after removal of the outlier, is indicated in Figure~\ref{fig:results_dist} and is consistent with the expected distribution $P(2\hcF)$.
We conclude that no continuous wave signals are detected.

\section{Sensitivity}

Table~\ref{tab:h0sens} lists the estimated sensitivity $h_0 = \num{1.01e-25}$ achieved by the search.
This is a statistical statement at 90\% confidence: were a large number of continuous wave signals present within the parameter space, with the given $h_0$ and other amplitude parameters chosen at random, we would have detected 90\% of them.
This statement is validated by performing 500 searches of small regions of the parameter space, each containing a simulated signal as described above, and confirming the expected 90\% detection rate.
For comparison with previous searches, we have also estimated the sensitivity at 95\% confidence, which gives $h_0 = \num{1.07e-25}$.

\section{Discussion}

Table~\ref{tab:h0sens} lists the sensitivity $h_0$ achieved by previous searches for continuous waves; they covered parameter spaces including, and significantly larger than, the parameter space of this search.
Our search improves in sensitivity by 24--69\% over previous searches, and is the most sensitive exploration of this parameter space yet performed.

While optimisation techniques are used to select the most sensitive continuous wave search, when the parameter space is specified \emph{a priori}~\cite{PrixShal2012:SCntGrvWOpStMFCmC,MessEtAl2015:GrvWSXCmpSMPrDtADt,WalsEtAl2016:CmpMtDtGrvWUnNS,WalsEtAl2019:OpCAnMASCnGrvWEns}, optimisation of the parameter space and search setup simultaneously is uncommon~\cite{MingEtAl2016:OpDrSrCntGrvW}.
Our approach realises a simple astrophysically-motivated benchmark [Eq.~\eqref{eq:benchmark}] as optimised choices for both parameter space and search setup.
It makes explicit the prior assumptions used to construct the parameter space, which then permits those assumptions to be refined by improved understanding of neutron star physics and Galactic neutron star populations.

The minimum spin-down $\fdmin$ (Table~\ref{tab:params}) is smaller than used in the previous searches listed in Table~\ref{tab:h0sens}.
While~\cite{DergPapa2020:RsFASCnGrvWSmlSr,dergachev2020results} also focus on small spin-downs, our choice of $\fdmin$ was not made \emph{a priori}, but arose as a consequence of Eq.~\eqref{eq:benchmark}.
The chosen $\fdmin$ and Eq.~\eqref{eq:dist} imply that the search is sensitive to gravitationally-radiating neutron stars within a distance $d = \SI{320}{pc}$.
Of the 289 pulsars indicated in Figure~\ref{fig:spindown_range}, only 10 are within this distance.
Nevertheless, only a small fraction ($\sim \num{3e3}/\num{e8}$) of Galactic neutron stars are observed as pulsars, and therefore one might naively expect $\sim \num{3e5}$ electromagnetically-quiet neutron stars within this distance.
A more pessimistic count may be derived starting from the modelled volume density of neutron stars in the solar neighbourhood~\cite{SartEtAl2010:GlcNtStISVlDstDH} of $\sim 1$--$5{\times}10^{-4} \si{pc^{-3}}$; this suggests only $\sim \num{3e3}$--$\num{2e5}$ neutron stars within a volume with radius $d = \SI{320}{pc}$.
Taken together with the estimated number of pulsars within the searched frequency band as a fraction of the number of observed pulsars ($\sim 0.2 / \num{3e3}$), the number of neutron stars within the searched band and sensitive volume might be $\sim 20$ in the naive estimate, or $\sim 0.2$--$13$ in the pessimistic estimate.
Given the many assumptions and simplicity of the above calculations, at best this suggests substantial uncertainty in the number of electromagnetically-quiet neutron stars this search is sensitive to.

Neutron star non-axisymmetry is characterised by the equatorial ellipticity $\epsilon$.
The minimum $\epsilon$ to which this search is sensitive at 90\% confidence and $d = \SI{320}{pc}$~\cite{BonaGour1996:GrvWPlEmMgFIDst},
\begin{equation}
    \label{eq:ellipsens}
    \epsilon = \frac{ c^4 h_0 d }{4 \pi^2 G \Izz f^2} = \num{1.04e-6} \,,
\end{equation}
is within conservative maximum values attainable by theoretical models~\cite{JohnOwen2013:MxElDfrRltSt}.
A detection of continuous waves by this search was certainly possible, therefore, based on current knowledge of neutron star physics.

The sensitivity achieved by this search confirms the advantages of a depth-first strategy for all-sky continuous wave surveys.
Such a strategy, in concert with complementary breadth-first surveys of wide parameter spaces, should continue to be pursued.
There is ample scope to refine the benchmark of Eq.~\eqref{eq:benchmark}, perhaps by including a more informed distribution of Galactic neutron stars spinning down through electromagnetic and gravitational waves.
An insightful choice of benchmark could be pivotal to a first detection of continuous gravitational waves.

\acknowledgements

We thank Hannah Middleton for helpful comments on the manuscript.
This research was supported by the Australian Research Council Centre of Excellence for Gravitational Wave Discovery (OzGrav) through project number CE170100004.
It used data, software and/or web tools obtained from the Gravitational Wave Open Science Center (\url{https://www.gw-openscience.org/}), a service of LIGO Laboratory, the LIGO Scientific Collaboration and the Virgo Collaboration. LIGO Laboratory and Advanced LIGO are funded by the United States National Science Foundation (NSF) as well as the Science and Technology Facilities Council (STFC) of the United Kingdom, the Max-Planck-Society (MPS), and the State of Niedersachsen/Germany for support of the construction of Advanced LIGO and construction and operation of the GEO600 detector. Additional support for Advanced LIGO was provided by the Australian Research Council. Virgo is funded, through the European Gravitational Observatory (EGO), by the French Centre National de Recherche Scientifique (CNRS), the Italian Istituto Nazionale della Fisica Nucleare (INFN) and the Dutch Nikhef, with contributions by institutions from Belgium, Germany, Greece, Hungary, Ireland, Japan, Monaco, Poland, Portugal, Spain.
It used the software packages LALSuite~\cite{lalsuite}, Octave~\cite{octave}, OctApps~\cite{WettEtAl2018:OcLOFnCntGrvDAn}, Python~\cite{python}, NumPy~\cite{numpy}, Matplotlib~\cite{matplotlib}, and SciPy~\cite{scipy}.
The search was performed on the OzSTAR national facility at Swinburne University of Technology. The OzSTAR program receives funding in part from the Astronomy National Collaborative Research Infrastructure Strategy (NCRIS) allocation provided by the Australian Government.
Document number LIGO-P2000536.


\begin{thebibliography}{69}%
\makeatletter
\providecommand \@ifxundefined [1]{%
 \@ifx{#1\undefined}
}%
\providecommand \@ifnum [1]{%
 \ifnum #1\expandafter \@firstoftwo
 \else \expandafter \@secondoftwo
 \fi
}%
\providecommand \@ifx [1]{%
 \ifx #1\expandafter \@firstoftwo
 \else \expandafter \@secondoftwo
 \fi
}%
\providecommand \natexlab [1]{#1}%
\providecommand \enquote  [1]{``#1''}%
\providecommand \bibnamefont  [1]{#1}%
\providecommand \bibfnamefont [1]{#1}%
\providecommand \citenamefont [1]{#1}%
\providecommand \href@noop [0]{\@secondoftwo}%
\providecommand \href [0]{\begingroup \@sanitize@url \@href}%
\providecommand \@href[1]{\@@startlink{#1}\@@href}%
\providecommand \@@href[1]{\endgroup#1\@@endlink}%
\providecommand \@sanitize@url [0]{\catcode `\\12\catcode `\$12\catcode
  `\&12\catcode `\#12\catcode `\^12\catcode `\_12\catcode `\%12\relax}%
\providecommand \@@startlink[1]{}%
\providecommand \@@endlink[0]{}%
\providecommand \url  [0]{\begingroup\@sanitize@url \@url }%
\providecommand \@url [1]{\endgroup\@href {#1}{\urlprefix }}%
\providecommand \urlprefix  [0]{URL }%
\providecommand \Eprint [0]{\href }%
\providecommand \doibase [0]{https://doi.org/}%
\providecommand \selectlanguage [0]{\@gobble}%
\providecommand \bibinfo  [0]{\@secondoftwo}%
\providecommand \bibfield  [0]{\@secondoftwo}%
\providecommand \translation [1]{[#1]}%
\providecommand \BibitemOpen [0]{}%
\providecommand \bibitemStop [0]{}%
\providecommand \bibitemNoStop [0]{.\EOS\space}%
\providecommand \EOS [0]{\spacefactor3000\relax}%
\providecommand \BibitemShut  [1]{\csname bibitem#1\endcsname}%
\let\auto@bib@innerbib\@empty
\bibitem [{\citenamefont {{Aasi}}\ \emph {et~al.}(2015)\citenamefont {{Aasi}}
  \emph {et~al.}}]{LIGO2015:AdvLIG}%
  \BibitemOpen
  \bibfield  {author} {\bibinfo {author} {\bibfnamefont {J.}~\bibnamefont
  {{Aasi}}} \emph {et~al.} (\bibinfo {collaboration} {{LIGO Scientific
  Collaboration}}),\ }\bibfield  {title} {\bibinfo {title} {{Advanced LIGO}},\
  }\href {https://doi.org/10.1088/0264-9381/32/7/074001} {\bibfield  {journal}
  {\bibinfo  {journal} {Classical and Quantum Gravity}\ }\textbf {\bibinfo
  {volume} {32}},\ \bibinfo {eid} {074001} (\bibinfo {year} {2015})},\ \Eprint
  {https://arxiv.org/abs/1411.4547} {arXiv:1411.4547 [gr-qc]} \BibitemShut
  {NoStop}%
\bibitem [{\citenamefont {{Acernese}}\ \emph {et~al.}(2018)\citenamefont
  {{Acernese}} \emph {et~al.}}]{AcerEtAl2018:SttAdvVr}%
  \BibitemOpen
  \bibfield  {author} {\bibinfo {author} {\bibfnamefont {F.}~\bibnamefont
  {{Acernese}}} \emph {et~al.},\ }\bibfield  {title} {\bibinfo {title} {{Status
  of Advanced Virgo}},\ }in\ \href
  {https://doi.org/10.1051/epjconf/201818202003} {\emph {\bibinfo {booktitle}
  {{European Physical Journal Web of Conferences}}}},\ Vol.\ \bibinfo {volume}
  {182}\ (\bibinfo {year} {2018})\ p.\ \bibinfo {pages} {02003}\BibitemShut
  {NoStop}%
\bibitem [{\citenamefont {{Abbott}}\ \emph
  {et~al.}(2017{\natexlab{a}})\citenamefont {{Abbott}} \emph
  {et~al.}}]{LIGOVirg2017:GWObsGrvWBNtSIn}%
  \BibitemOpen
  \bibfield  {author} {\bibinfo {author} {\bibfnamefont {B.~P.}\ \bibnamefont
  {{Abbott}}} \emph {et~al.} (\bibinfo {collaboration} {{LIGO Scientific
  Collaboration} and {Virgo Collaboration}}),\ }\bibfield  {title} {\bibinfo
  {title} {{GW170817: Observation of Gravitational Waves from a Binary Neutron
  Star Inspiral}},\ }\href {https://doi.org/10.1103/PhysRevLett.119.161101}
  {\bibfield  {journal} {\bibinfo  {journal} {Physical Review Letters}\
  }\textbf {\bibinfo {volume} {119}},\ \bibinfo {eid} {161101} (\bibinfo {year}
  {2017}{\natexlab{a}})},\ \Eprint {https://arxiv.org/abs/1710.05832}
  {arXiv:1710.05832 [gr-qc]} \BibitemShut {NoStop}%
\bibitem [{\citenamefont {{Abbott}}\ \emph
  {et~al.}(2017{\natexlab{b}})\citenamefont {{Abbott}} \emph
  {et~al.}}]{LIGOVirg2017:MltObsBnNtSMr}%
  \BibitemOpen
  \bibfield  {author} {\bibinfo {author} {\bibfnamefont {B.~P.}\ \bibnamefont
  {{Abbott}}} \emph {et~al.} (\bibinfo {collaboration} {{LIGO Scientific
  Collaboration} and {Virgo Collaboration} and others}),\ }\bibfield  {title}
  {\bibinfo {title} {{Multi-messenger Observations of a Binary Neutron Star
  Merger}},\ }\href {https://doi.org/10.3847/2041-8213/aa91c9} {\bibfield
  {journal} {\bibinfo  {journal} {Astrophysical Journal}\ }\textbf {\bibinfo
  {volume} {848}},\ \bibinfo {eid} {L12} (\bibinfo {year}
  {2017}{\natexlab{b}})},\ \Eprint {https://arxiv.org/abs/1710.05833}
  {arXiv:1710.05833 [astro-ph.HE]} \BibitemShut {NoStop}%
\bibitem [{\citenamefont {{Riles}}(2017)}]{Rile2017:RcSrCntGrvWv}%
  \BibitemOpen
  \bibfield  {author} {\bibinfo {author} {\bibfnamefont {K.}~\bibnamefont
  {{Riles}}},\ }\bibfield  {title} {\bibinfo {title} {{Recent searches for
  continuous gravitational waves}},\ }\href
  {https://doi.org/10.1142/S021773231730035X} {\bibfield  {journal} {\bibinfo
  {journal} {Modern Physics Letters A}\ }\textbf {\bibinfo {volume} {32}},\
  \bibinfo {eid} {1730035-685} (\bibinfo {year} {2017})},\ \Eprint
  {https://arxiv.org/abs/1712.05897} {arXiv:1712.05897 [gr-qc]} \BibitemShut
  {NoStop}%
\bibitem [{\citenamefont {{Sieniawska}}\ and\ \citenamefont
  {{Bejger}}(2019)}]{SienBejg2019:CntGrvWNtSCrSPr}%
  \BibitemOpen
  \bibfield  {author} {\bibinfo {author} {\bibfnamefont {M.}~\bibnamefont
  {{Sieniawska}}}\ and\ \bibinfo {author} {\bibfnamefont {M.}~\bibnamefont
  {{Bejger}}},\ }\bibfield  {title} {\bibinfo {title} {{Continuous
  Gravitational Waves from Neutron Stars: Current Status and Prospects}},\
  }\href {https://doi.org/10.3390/universe5110217} {\bibfield  {journal}
  {\bibinfo  {journal} {Universe}\ }\textbf {\bibinfo {volume} {5}},\ \bibinfo
  {pages} {217} (\bibinfo {year} {2019})},\ \Eprint
  {https://arxiv.org/abs/1909.12600} {arXiv:1909.12600 [astro-ph.HE]}
  \BibitemShut {NoStop}%
\bibitem [{\citenamefont {{Melatos}}\ and\ \citenamefont
  {{Payne}}(2005)}]{MelaPayn2005:GrvRdAcMlPMgnCM}%
  \BibitemOpen
  \bibfield  {author} {\bibinfo {author} {\bibfnamefont {A.}~\bibnamefont
  {{Melatos}}}\ and\ \bibinfo {author} {\bibfnamefont {D.~J.~B.}\ \bibnamefont
  {{Payne}}},\ }\bibfield  {title} {\bibinfo {title} {{Gravitational Radiation
  from an Accreting Millisecond Pulsar with a Magnetically Confined
  Mountain}},\ }\href {https://doi.org/10.1086/428600} {\bibfield  {journal}
  {\bibinfo  {journal} {Astrophysical Journal}\ }\textbf {\bibinfo {volume}
  {623}},\ \bibinfo {pages} {1044} (\bibinfo {year} {2005})},\ \Eprint
  {https://arxiv.org/abs/astro-ph/0503287} {arXiv:astro-ph/0503287}
  \BibitemShut {NoStop}%
\bibitem [{\citenamefont {{Mastrano}}\ and\ \citenamefont
  {{Melatos}}(2012)}]{MastMela2012:UGrvULIntMgFStRcP}%
  \BibitemOpen
  \bibfield  {author} {\bibinfo {author} {\bibfnamefont {A.}~\bibnamefont
  {{Mastrano}}}\ and\ \bibinfo {author} {\bibfnamefont {A.}~\bibnamefont
  {{Melatos}}},\ }\bibfield  {title} {\bibinfo {title} {{Updated
  gravitational-wave upper limits on the internal magnetic field strength of
  recycled pulsars}},\ }\href
  {https://doi.org/10.1111/j.1365-2966.2011.20350.x} {\bibfield  {journal}
  {\bibinfo  {journal} {Monthly Notices of the Royal Astronomical Society}\
  }\textbf {\bibinfo {volume} {421}},\ \bibinfo {pages} {760} (\bibinfo {year}
  {2012})},\ \Eprint {https://arxiv.org/abs/1112.1542} {arXiv:1112.1542
  [astro-ph.HE]} \BibitemShut {NoStop}%
\bibitem [{\citenamefont {{Johnson-McDaniel}}\ and\ \citenamefont
  {{Owen}}(2013)}]{JohnOwen2013:MxElDfrRltSt}%
  \BibitemOpen
  \bibfield  {author} {\bibinfo {author} {\bibfnamefont {N.~K.}\ \bibnamefont
  {{Johnson-McDaniel}}}\ and\ \bibinfo {author} {\bibfnamefont {B.~J.}\
  \bibnamefont {{Owen}}},\ }\bibfield  {title} {\bibinfo {title} {{Maximum
  elastic deformations of relativistic stars}},\ }\href
  {https://doi.org/10.1103/PhysRevD.88.044004} {\bibfield  {journal} {\bibinfo
  {journal} {Physical Review D}\ }\textbf {\bibinfo {volume} {88}},\ \bibinfo
  {eid} {044004} (\bibinfo {year} {2013})},\ \Eprint
  {https://arxiv.org/abs/1208.5227} {arXiv:1208.5227 [astro-ph.SR]}
  \BibitemShut {NoStop}%
\bibitem [{\citenamefont {{Woan}}\ \emph {et~al.}(2018)\citenamefont {{Woan}},
  \citenamefont {{Pitkin}}, \citenamefont {{Haskell}}, \citenamefont
  {{Jones}},\ and\ \citenamefont {{Lasky}}}]{WoanEtAl2018:EvMnEllMllPl}%
  \BibitemOpen
  \bibfield  {author} {\bibinfo {author} {\bibfnamefont {G.}~\bibnamefont
  {{Woan}}}, \bibinfo {author} {\bibfnamefont {M.~D.}\ \bibnamefont
  {{Pitkin}}}, \bibinfo {author} {\bibfnamefont {B.}~\bibnamefont {{Haskell}}},
  \bibinfo {author} {\bibfnamefont {D.~I.}\ \bibnamefont {{Jones}}},\ and\
  \bibinfo {author} {\bibfnamefont {P.~D.}\ \bibnamefont {{Lasky}}},\
  }\bibfield  {title} {\bibinfo {title} {{Evidence for a Minimum Ellipticity in
  Millisecond Pulsars}},\ }\href {https://doi.org/10.3847/2041-8213/aad86a}
  {\bibfield  {journal} {\bibinfo  {journal} {Astrophysical Journal}\ }\textbf
  {\bibinfo {volume} {863}},\ \bibinfo {eid} {L40} (\bibinfo {year} {2018})},\
  \Eprint {https://arxiv.org/abs/1806.02822} {arXiv:1806.02822 [astro-ph.HE]}
  \BibitemShut {NoStop}%
\bibitem [{\citenamefont {{Gittins}}\ \emph {et~al.}(2021)\citenamefont
  {{Gittins}}, \citenamefont {{Andersson}},\ and\ \citenamefont
  {{Jones}}}]{gittins2020modelling}%
  \BibitemOpen
  \bibfield  {author} {\bibinfo {author} {\bibfnamefont {F.}~\bibnamefont
  {{Gittins}}}, \bibinfo {author} {\bibfnamefont {N.}~\bibnamefont
  {{Andersson}}},\ and\ \bibinfo {author} {\bibfnamefont {D.~I.}\ \bibnamefont
  {{Jones}}},\ }\bibfield  {title} {\bibinfo {title} {{Modelling neutron star
  mountains}},\ }\href {https://doi.org/10.1093/mnras/staa3635} {\bibfield
  {journal} {\bibinfo  {journal} {Monthly Notices of the Royal Astronomical
  Society}\ }\textbf {\bibinfo {volume} {500}},\ \bibinfo {pages} {5570}
  (\bibinfo {year} {2021})},\ \Eprint {https://arxiv.org/abs/2009.12794}
  {arXiv:2009.12794 [astro-ph.HE]} \BibitemShut {NoStop}%
\bibitem [{\citenamefont {{Palomba}}(2005)}]{Palo2005:SmlPpIsNSEvTEGrvW}%
  \BibitemOpen
  \bibfield  {author} {\bibinfo {author} {\bibfnamefont {C.}~\bibnamefont
  {{Palomba}}},\ }\bibfield  {title} {\bibinfo {title} {{Simulation of a
  population of isolated neutron stars evolving through the emission of
  gravitational waves}},\ }\href
  {https://doi.org/10.1111/j.1365-2966.2005.08975.x} {\bibfield  {journal}
  {\bibinfo  {journal} {Monthly Notices of the Royal Astronomical Society}\
  }\textbf {\bibinfo {volume} {359}},\ \bibinfo {pages} {1150} (\bibinfo {year}
  {2005})},\ \Eprint {https://arxiv.org/abs/astro-ph/0503046}
  {arXiv:astro-ph/0503046} \BibitemShut {NoStop}%
\bibitem [{\citenamefont {{Wade}}\ \emph {et~al.}(2012)\citenamefont {{Wade}},
  \citenamefont {{Siemens}}, \citenamefont {{Kaplan}}, \citenamefont
  {{Knispel}},\ and\ \citenamefont {{Allen}}}]{WadeEtAl2012:CntGrvWIsGlNSAdDE}%
  \BibitemOpen
  \bibfield  {author} {\bibinfo {author} {\bibfnamefont {L.}~\bibnamefont
  {{Wade}}}, \bibinfo {author} {\bibfnamefont {X.}~\bibnamefont {{Siemens}}},
  \bibinfo {author} {\bibfnamefont {D.~L.}\ \bibnamefont {{Kaplan}}}, \bibinfo
  {author} {\bibfnamefont {B.}~\bibnamefont {{Knispel}}},\ and\ \bibinfo
  {author} {\bibfnamefont {B.}~\bibnamefont {{Allen}}},\ }\bibfield  {title}
  {\bibinfo {title} {{Continuous gravitational waves from isolated Galactic
  neutron stars in the advanced detector era}},\ }\href
  {https://doi.org/10.1103/PhysRevD.86.124011} {\bibfield  {journal} {\bibinfo
  {journal} {Physical Review D}\ }\textbf {\bibinfo {volume} {86}},\ \bibinfo
  {eid} {124011} (\bibinfo {year} {2012})},\ \Eprint
  {https://arxiv.org/abs/1209.2971} {arXiv:1209.2971 [gr-qc]} \BibitemShut
  {NoStop}%
\bibitem [{\citenamefont {{Brady}}\ \emph {et~al.}(1998)\citenamefont
  {{Brady}}, \citenamefont {{Creighton}}, \citenamefont {{Cutler}},\ and\
  \citenamefont {{Schutz}}}]{BradEtAl1998:SrcPrdSrLI}%
  \BibitemOpen
  \bibfield  {author} {\bibinfo {author} {\bibfnamefont {P.~R.}\ \bibnamefont
  {{Brady}}}, \bibinfo {author} {\bibfnamefont {T.}~\bibnamefont
  {{Creighton}}}, \bibinfo {author} {\bibfnamefont {C.}~\bibnamefont
  {{Cutler}}},\ and\ \bibinfo {author} {\bibfnamefont {B.~F.}\ \bibnamefont
  {{Schutz}}},\ }\bibfield  {title} {\bibinfo {title} {{Searching for periodic
  sources with LIGO}},\ }\href {https://doi.org/10.1103/PhysRevD.57.2101}
  {\bibfield  {journal} {\bibinfo  {journal} {Physical Review D}\ }\textbf
  {\bibinfo {volume} {57}},\ \bibinfo {pages} {2101} (\bibinfo {year}
  {1998})},\ \Eprint {https://arxiv.org/abs/gr-qc/9702050}
  {arXiv:gr-qc/9702050} \BibitemShut {NoStop}%
\bibitem [{\citenamefont {{Abbott}}\ \emph
  {et~al.}(2019{\natexlab{a}})\citenamefont {{Abbott}} \emph
  {et~al.}}]{LIGOVirg2019:SrGrvWvKPlTHrm20152017LD}%
  \BibitemOpen
  \bibfield  {author} {\bibinfo {author} {\bibfnamefont {B.~P.}\ \bibnamefont
  {{Abbott}}} \emph {et~al.} (\bibinfo {collaboration} {{LIGO Scientific
  Collaboration} and {Virgo Collaboration} and others}),\ }\bibfield  {title}
  {\bibinfo {title} {{Searches for Gravitational Waves from Known Pulsars at
  Two Harmonics in 2015-2017 LIGO Data}},\ }\href
  {https://doi.org/10.3847/1538-4357/ab20cb} {\bibfield  {journal} {\bibinfo
  {journal} {Astrophysical Journal}\ }\textbf {\bibinfo {volume} {879}},\
  \bibinfo {eid} {10} (\bibinfo {year} {2019}{\natexlab{a}})},\ \Eprint
  {https://arxiv.org/abs/1902.08507} {arXiv:1902.08507 [astro-ph.HE]}
  \BibitemShut {NoStop}%
\bibitem [{\citenamefont {{Abbott}}\ \emph
  {et~al.}(2019{\natexlab{b}})\citenamefont {{Abbott}} \emph
  {et~al.}}]{LIGOVirg2019:NrrSrGrvWKPlUSLObR}%
  \BibitemOpen
  \bibfield  {author} {\bibinfo {author} {\bibfnamefont {B.~P.}\ \bibnamefont
  {{Abbott}}} \emph {et~al.} (\bibinfo {collaboration} {{LIGO Scientific
  Collaboration} and {Virgo Collaboration} and others}),\ }\bibfield  {title}
  {\bibinfo {title} {{Narrow-band search for gravitational waves from known
  pulsars using the second LIGO observing run}},\ }\href
  {https://doi.org/10.1103/PhysRevD.99.122002} {\bibfield  {journal} {\bibinfo
  {journal} {Physical Review D}\ }\textbf {\bibinfo {volume} {99}},\ \bibinfo
  {eid} {122002} (\bibinfo {year} {2019}{\natexlab{b}})}\BibitemShut {NoStop}%
\bibitem [{\citenamefont {{Abbott}}\ \emph {et~al.}(2020)\citenamefont
  {{Abbott}} \emph {et~al.}}]{LIGOVirg2020:GrvCnsEqElMlP}%
  \BibitemOpen
  \bibfield  {author} {\bibinfo {author} {\bibfnamefont {R.}~\bibnamefont
  {{Abbott}}} \emph {et~al.} (\bibinfo {collaboration} {{LIGO Scientific
  Collaboration} and {Virgo Collaboration} and others}),\ }\bibfield  {title}
  {\bibinfo {title} {{Gravitational-wave Constraints on the Equatorial
  Ellipticity of Millisecond Pulsars}},\ }\href
  {https://doi.org/10.3847/2041-8213/abb655} {\bibfield  {journal} {\bibinfo
  {journal} {Astrophysical Journal}\ }\textbf {\bibinfo {volume} {902}},\
  \bibinfo {pages} {L21} (\bibinfo {year} {2020})}\BibitemShut {NoStop}%
\bibitem [{\citenamefont {{Jones}}\ and\ \citenamefont
  {{Andersson}}(2002)}]{JoneAnde2002:GrvWvFrPrcNtS}%
  \BibitemOpen
  \bibfield  {author} {\bibinfo {author} {\bibfnamefont {D.~I.}\ \bibnamefont
  {{Jones}}}\ and\ \bibinfo {author} {\bibfnamefont {N.}~\bibnamefont
  {{Andersson}}},\ }\bibfield  {title} {\bibinfo {title} {{Gravitational waves
  from freely precessing neutron stars}},\ }\href
  {https://doi.org/10.1046/j.1365-8711.2002.05180.x} {\bibfield  {journal}
  {\bibinfo  {journal} {Monthly Notices of the Royal Astronomical Society}\
  }\textbf {\bibinfo {volume} {331}},\ \bibinfo {pages} {203} (\bibinfo {year}
  {2002})},\ \Eprint {https://arxiv.org/abs/gr-qc/0106094}
  {arXiv:gr-qc/0106094} \BibitemShut {NoStop}%
\bibitem [{\citenamefont {{Abbott}}\ \emph
  {et~al.}(2019{\natexlab{c}})\citenamefont {{Abbott}} \emph
  {et~al.}}]{LIGOVirg2019:ASCntGrvWIsNtSUAdLOD}%
  \BibitemOpen
  \bibfield  {author} {\bibinfo {author} {\bibfnamefont {B.~P.}\ \bibnamefont
  {{Abbott}}} \emph {et~al.} (\bibinfo {collaboration} {{LIGO Scientific
  Collaboration} and {Virgo Collaboration} and others}),\ }\bibfield  {title}
  {\bibinfo {title} {{All-sky search for continuous gravitational waves from
  isolated neutron stars using Advanced LIGO O2 data}},\ }\href
  {https://doi.org/10.1103/PhysRevD.100.024004} {\bibfield  {journal} {\bibinfo
   {journal} {Physical Review D}\ }\textbf {\bibinfo {volume} {100}},\ \bibinfo
  {eid} {024004} (\bibinfo {year} {2019}{\natexlab{c}})},\ \Eprint
  {https://arxiv.org/abs/1903.01901} {arXiv:1903.01901 [astro-ph.HE]}
  \BibitemShut {NoStop}%
\bibitem [{\citenamefont {{Steltner}}\ \emph {et~al.}(2021)\citenamefont
  {{Steltner}}, \citenamefont {{Papa}}, \citenamefont {{Eggenstein}},
  \citenamefont {{Allen}}, \citenamefont {{Dergachev}}, \citenamefont {{Prix}},
  \citenamefont {{Machenschalk}}, \citenamefont {{Walsh}}, \citenamefont
  {{Zhu}}, \citenamefont {{Behnke}},\ and\ \citenamefont
  {{Kwang}}}]{O2-EinsteinAtHome}%
  \BibitemOpen
  \bibfield  {author} {\bibinfo {author} {\bibfnamefont {B.}~\bibnamefont
  {{Steltner}}}, \bibinfo {author} {\bibfnamefont {M.~A.}\ \bibnamefont
  {{Papa}}}, \bibinfo {author} {\bibfnamefont {H.-B.}\ \bibnamefont
  {{Eggenstein}}}, \bibinfo {author} {\bibfnamefont {B.}~\bibnamefont
  {{Allen}}}, \bibinfo {author} {\bibfnamefont {V.}~\bibnamefont
  {{Dergachev}}}, \bibinfo {author} {\bibfnamefont {R.}~\bibnamefont {{Prix}}},
  \bibinfo {author} {\bibfnamefont {B.}~\bibnamefont {{Machenschalk}}},
  \bibinfo {author} {\bibfnamefont {S.}~\bibnamefont {{Walsh}}}, \bibinfo
  {author} {\bibfnamefont {S.~J.}\ \bibnamefont {{Zhu}}}, \bibinfo {author}
  {\bibfnamefont {O.}~\bibnamefont {{Behnke}}},\ and\ \bibinfo {author}
  {\bibfnamefont {S.}~\bibnamefont {{Kwang}}},\ }\bibfield  {title} {\bibinfo
  {title} {{Einstein@Home All-sky Search for Continuous Gravitational Waves in
  LIGO O2 Public Data}},\ }\href {https://doi.org/10.3847/1538-4357/abc7c9}
  {\bibfield  {journal} {\bibinfo  {journal} {Astrophysical Journal}\ }\textbf
  {\bibinfo {volume} {909}},\ \bibinfo {eid} {79} (\bibinfo {year} {2021})},\
  \Eprint {https://arxiv.org/abs/2009.12260} {arXiv:2009.12260 [astro-ph.HE]}
  \BibitemShut {NoStop}%
\bibitem [{\citenamefont {{Covas}}\ and\ \citenamefont
  {{Sintes}}(2020)}]{CovaSint2020:FASCntGrvSgUnNSBSUAdLD}%
  \BibitemOpen
  \bibfield  {author} {\bibinfo {author} {\bibfnamefont {P.~B.}\ \bibnamefont
  {{Covas}}}\ and\ \bibinfo {author} {\bibfnamefont {A.~M.}\ \bibnamefont
  {{Sintes}}},\ }\bibfield  {title} {\bibinfo {title} {{First All-Sky Search
  for Continuous Gravitational-Wave Signals from Unknown Neutron Stars in
  Binary Systems Using Advanced LIGO Data}},\ }\href
  {https://doi.org/10.1103/PhysRevLett.124.191102} {\bibfield  {journal}
  {\bibinfo  {journal} {Physical Review Letters}\ }\textbf {\bibinfo {volume}
  {124}},\ \bibinfo {eid} {191102} (\bibinfo {year} {2020})},\ \Eprint
  {https://arxiv.org/abs/2001.08411} {arXiv:2001.08411 [gr-qc]} \BibitemShut
  {NoStop}%
\bibitem [{\citenamefont {{Abbott}}\ \emph
  {et~al.}(2019{\natexlab{d}})\citenamefont {{Abbott}} \emph
  {et~al.}}]{LIGOVirg2019:SrCntGrvW15SpRFmBAL}%
  \BibitemOpen
  \bibfield  {author} {\bibinfo {author} {\bibfnamefont {B.~P.}\ \bibnamefont
  {{Abbott}}} \emph {et~al.} (\bibinfo {collaboration} {{LIGO Scientific
  Collaboration} and {Virgo Collaboration}}),\ }\bibfield  {title} {\bibinfo
  {title} {{Searches for Continuous Gravitational Waves from 15 Supernova
  Remnants and Fomalhaut b with Advanced LIGO}},\ }\href
  {https://doi.org/10.3847/1538-4357/ab113b} {\bibfield  {journal} {\bibinfo
  {journal} {Astrophysical Journal}\ }\textbf {\bibinfo {volume} {875}},\
  \bibinfo {eid} {122} (\bibinfo {year} {2019}{\natexlab{d}})}\BibitemShut
  {NoStop}%
\bibitem [{\citenamefont {{Lindblom}}\ and\ \citenamefont
  {{Owen}}(2020)}]{LindOwen2020:DrSCntGrvWTSpRDALSObR}%
  \BibitemOpen
  \bibfield  {author} {\bibinfo {author} {\bibfnamefont {L.}~\bibnamefont
  {{Lindblom}}}\ and\ \bibinfo {author} {\bibfnamefont {B.~J.}\ \bibnamefont
  {{Owen}}},\ }\bibfield  {title} {\bibinfo {title} {{Directed searches for
  continuous gravitational waves from twelve supernova remnants in data from
  Advanced LIGO's second observing run}},\ }\href
  {https://doi.org/10.1103/PhysRevD.101.083023} {\bibfield  {journal} {\bibinfo
   {journal} {Physical Review D}\ }\textbf {\bibinfo {volume} {101}},\ \bibinfo
  {eid} {083023} (\bibinfo {year} {2020})},\ \Eprint
  {https://arxiv.org/abs/2003.00072} {arXiv:2003.00072 [gr-qc]} \BibitemShut
  {NoStop}%
\bibitem [{\citenamefont {{Millhouse}}\ \emph {et~al.}(2020)\citenamefont
  {{Millhouse}}, \citenamefont {{Strang}},\ and\ \citenamefont
  {{Melatos}}}]{MillEtAl2020:SGrvW12YSprRmHMMAdLSObR}%
  \BibitemOpen
  \bibfield  {author} {\bibinfo {author} {\bibfnamefont {M.}~\bibnamefont
  {{Millhouse}}}, \bibinfo {author} {\bibfnamefont {L.}~\bibnamefont
  {{Strang}}},\ and\ \bibinfo {author} {\bibfnamefont {A.}~\bibnamefont
  {{Melatos}}},\ }\bibfield  {title} {\bibinfo {title} {{Search for
  gravitational waves from 12 young supernova remnants with a hidden Markov
  model in Advanced LIGO's second observing run}},\ }\href
  {https://doi.org/10.1103/PhysRevD.102.083025} {\bibfield  {journal} {\bibinfo
   {journal} {Physical Review D}\ }\textbf {\bibinfo {volume} {102}},\ \bibinfo
  {eid} {083025} (\bibinfo {year} {2020})},\ \Eprint
  {https://arxiv.org/abs/2003.08588} {arXiv:2003.08588 [gr-qc]} \BibitemShut
  {NoStop}%
\bibitem [{\citenamefont {{Papa}}\ \emph {et~al.}(2020)\citenamefont {{Papa}},
  \citenamefont {{Ming}}, \citenamefont {{Gotthelf}}, \citenamefont {{Allen}},
  \citenamefont {{Prix}}, \citenamefont {{Dergachev}}, \citenamefont
  {{Eggenstein}}, \citenamefont {{Singh}},\ and\ \citenamefont
  {{Zhu}}}]{PapaEtAl2020:SCntGrvWCCOSpRmCsVJG}%
  \BibitemOpen
  \bibfield  {author} {\bibinfo {author} {\bibfnamefont {M.~A.}\ \bibnamefont
  {{Papa}}}, \bibinfo {author} {\bibfnamefont {J.}~\bibnamefont {{Ming}}},
  \bibinfo {author} {\bibfnamefont {E.~V.}\ \bibnamefont {{Gotthelf}}},
  \bibinfo {author} {\bibfnamefont {B.}~\bibnamefont {{Allen}}}, \bibinfo
  {author} {\bibfnamefont {R.}~\bibnamefont {{Prix}}}, \bibinfo {author}
  {\bibfnamefont {V.}~\bibnamefont {{Dergachev}}}, \bibinfo {author}
  {\bibfnamefont {H.-B.}\ \bibnamefont {{Eggenstein}}}, \bibinfo {author}
  {\bibfnamefont {A.}~\bibnamefont {{Singh}}},\ and\ \bibinfo {author}
  {\bibfnamefont {S.~J.}\ \bibnamefont {{Zhu}}},\ }\bibfield  {title} {\bibinfo
  {title} {{Search for Continuous Gravitational Waves from the Central Compact
  Objects in Supernova Remnants Cassiopeia A, Vela Jr., and G347.3-0.5}},\
  }\href {https://doi.org/10.3847/1538-4357/ab92a6} {\bibfield  {journal}
  {\bibinfo  {journal} {Astrophysical Journal}\ }\textbf {\bibinfo {volume}
  {897}},\ \bibinfo {eid} {22} (\bibinfo {year} {2020})},\ \Eprint
  {https://arxiv.org/abs/2005.06544} {arXiv:2005.06544 [astro-ph.HE]}
  \BibitemShut {NoStop}%
\bibitem [{\citenamefont {{Abbott}}\ \emph
  {et~al.}(2017{\natexlab{c}})\citenamefont {{Abbott}} \emph
  {et~al.}}]{LIGOVirg2017:ULGrvWScXMdCrsSAdLD}%
  \BibitemOpen
  \bibfield  {author} {\bibinfo {author} {\bibfnamefont {B.~P.}\ \bibnamefont
  {{Abbott}}} \emph {et~al.} (\bibinfo {collaboration} {{LIGO Scientific
  Collaboration} and {Virgo Collaboration} and others}),\ }\bibfield  {title}
  {\bibinfo {title} {{Upper Limits on Gravitational Waves from Scorpius X-1
  from a Model-based Cross-correlation Search in Advanced LIGO Data}},\ }\href
  {https://doi.org/10.3847/1538-4357/aa86f0} {\bibfield  {journal} {\bibinfo
  {journal} {Astrophysical Journal}\ }\textbf {\bibinfo {volume} {847}},\
  \bibinfo {eid} {47} (\bibinfo {year} {2017}{\natexlab{c}})}\BibitemShut
  {NoStop}%
\bibitem [{\citenamefont {{Abbott}}\ \emph
  {et~al.}(2019{\natexlab{e}})\citenamefont {{Abbott}} \emph
  {et~al.}}]{LIGOVirg2019:SGrvWScXSAdLObsRImHMM}%
  \BibitemOpen
  \bibfield  {author} {\bibinfo {author} {\bibfnamefont {B.~P.}\ \bibnamefont
  {{Abbott}}} \emph {et~al.} (\bibinfo {collaboration} {{LIGO Scientific
  Collaboration} and {Virgo Collaboration} and others}),\ }\bibfield  {title}
  {\bibinfo {title} {{Search for gravitational waves from Scorpius X-1 in the
  second Advanced LIGO observing run with an improved hidden Markov model}},\
  }\href {https://doi.org/10.1103/PhysRevD.100.122002} {\bibfield  {journal}
  {\bibinfo  {journal} {Physical Review D}\ }\textbf {\bibinfo {volume}
  {100}},\ \bibinfo {pages} {122002} (\bibinfo {year}
  {2019}{\natexlab{e}})}\BibitemShut {NoStop}%
\bibitem [{\citenamefont {{Middleton}}\ \emph {et~al.}(2020)\citenamefont
  {{Middleton}}, \citenamefont {{Clearwater}}, \citenamefont {{Melatos}},\ and\
  \citenamefont {{Dunn}}}]{MiddEtAl2020:SGrvWFLMXBnSAdLObsRImHMM}%
  \BibitemOpen
  \bibfield  {author} {\bibinfo {author} {\bibfnamefont {H.}~\bibnamefont
  {{Middleton}}}, \bibinfo {author} {\bibfnamefont {P.}~\bibnamefont
  {{Clearwater}}}, \bibinfo {author} {\bibfnamefont {A.}~\bibnamefont
  {{Melatos}}},\ and\ \bibinfo {author} {\bibfnamefont {L.}~\bibnamefont
  {{Dunn}}},\ }\bibfield  {title} {\bibinfo {title} {{Search for gravitational
  waves from five low mass x-ray binaries in the second Advanced LIGO observing
  run with an improved hidden Markov model}},\ }\href
  {https://doi.org/10.1103/PhysRevD.102.023006} {\bibfield  {journal} {\bibinfo
   {journal} {Physical Review D}\ }\textbf {\bibinfo {volume} {102}},\ \bibinfo
  {eid} {023006} (\bibinfo {year} {2020})},\ \Eprint
  {https://arxiv.org/abs/2006.06907} {arXiv:2006.06907 [astro-ph.HE]}
  \BibitemShut {NoStop}%
\bibitem [{\citenamefont {{Zhang}}\ \emph {et~al.}(2021)\citenamefont
  {{Zhang}}, \citenamefont {{Papa}}, \citenamefont {{Krishnan}},\ and\
  \citenamefont {{Watts}}}]{zhang2020search}%
  \BibitemOpen
  \bibfield  {author} {\bibinfo {author} {\bibfnamefont {Y.}~\bibnamefont
  {{Zhang}}}, \bibinfo {author} {\bibfnamefont {M.~A.}\ \bibnamefont {{Papa}}},
  \bibinfo {author} {\bibfnamefont {B.}~\bibnamefont {{Krishnan}}},\ and\
  \bibinfo {author} {\bibfnamefont {A.~L.}\ \bibnamefont {{Watts}}},\
  }\bibfield  {title} {\bibinfo {title} {{Search for Continuous Gravitational
  Waves from Scorpius X-1 in LIGO O2 Data}},\ }\href
  {https://doi.org/10.3847/2041-8213/abd256} {\bibfield  {journal} {\bibinfo
  {journal} {Astrophysical Journal Letters}\ }\textbf {\bibinfo {volume}
  {906}},\ \bibinfo {eid} {L14} (\bibinfo {year} {2021})},\ \Eprint
  {https://arxiv.org/abs/2011.04414} {arXiv:2011.04414 [astro-ph.HE]}
  \BibitemShut {NoStop}%
\bibitem [{\citenamefont {{Piccinni}}\ \emph {et~al.}(2020)\citenamefont
  {{Piccinni}}, \citenamefont {{Astone}}, \citenamefont {{D'Antonio}},
  \citenamefont {{Frasca}}, \citenamefont {{Intini}}, \citenamefont {{La
  Rosa}}, \citenamefont {{Leaci}}, \citenamefont {{Mastrogiovanni}},
  \citenamefont {{Miller}},\ and\ \citenamefont
  {{Palomba}}}]{PiccEtAl2020:DrSCntGrvSGlCALSObR}%
  \BibitemOpen
  \bibfield  {author} {\bibinfo {author} {\bibfnamefont {O.~J.}\ \bibnamefont
  {{Piccinni}}}, \bibinfo {author} {\bibfnamefont {P.}~\bibnamefont
  {{Astone}}}, \bibinfo {author} {\bibfnamefont {S.}~\bibnamefont
  {{D'Antonio}}}, \bibinfo {author} {\bibfnamefont {S.}~\bibnamefont
  {{Frasca}}}, \bibinfo {author} {\bibfnamefont {G.}~\bibnamefont {{Intini}}},
  \bibinfo {author} {\bibfnamefont {I.}~\bibnamefont {{La Rosa}}}, \bibinfo
  {author} {\bibfnamefont {P.}~\bibnamefont {{Leaci}}}, \bibinfo {author}
  {\bibfnamefont {S.}~\bibnamefont {{Mastrogiovanni}}}, \bibinfo {author}
  {\bibfnamefont {A.}~\bibnamefont {{Miller}}},\ and\ \bibinfo {author}
  {\bibfnamefont {C.}~\bibnamefont {{Palomba}}},\ }\bibfield  {title} {\bibinfo
  {title} {{Directed search for continuous gravitational-wave signals from the
  Galactic Center in the Advanced LIGO second observing run}},\ }\href
  {https://doi.org/10.1103/PhysRevD.101.082004} {\bibfield  {journal} {\bibinfo
   {journal} {Physical Review D}\ }\textbf {\bibinfo {volume} {101}},\ \bibinfo
  {eid} {082004} (\bibinfo {year} {2020})},\ \Eprint
  {https://arxiv.org/abs/1910.05097} {arXiv:1910.05097 [gr-qc]} \BibitemShut
  {NoStop}%
\bibitem [{\citenamefont {{Abbott}}\ \emph {et~al.}(2021)\citenamefont
  {{Abbott}} \emph {et~al.}}]{O2-data}%
  \BibitemOpen
  \bibfield  {author} {\bibinfo {author} {\bibfnamefont {R.}~\bibnamefont
  {{Abbott}}} \emph {et~al.} (\bibinfo {collaboration} {{LIGO Scientific
  Collaboration} and {Virgo Collaboration}}),\ }\bibfield  {title} {\bibinfo
  {title} {{Open data from the first and second observing runs of Advanced LIGO
  and Advanced Virgo}},\ }\href {https://doi.org/10.1016/j.softx.2021.100658}
  {\bibfield  {journal} {\bibinfo  {journal} {SoftwareX}\ }\textbf {\bibinfo
  {volume} {13}},\ \bibinfo {eid} {100658} (\bibinfo {year}
  {2021})}\BibitemShut {NoStop}%
\bibitem [{\citenamefont {{Jaranowski}}\ \emph {et~al.}(1998)\citenamefont
  {{Jaranowski}}, \citenamefont {{Kr{\'o}lak}},\ and\ \citenamefont
  {{Schutz}}}]{JaraEtAl1998:DAnGrvSgSpNSSDtc}%
  \BibitemOpen
  \bibfield  {author} {\bibinfo {author} {\bibfnamefont {P.}~\bibnamefont
  {{Jaranowski}}}, \bibinfo {author} {\bibfnamefont {A.}~\bibnamefont
  {{Kr{\'o}lak}}},\ and\ \bibinfo {author} {\bibfnamefont {B.~F.}\ \bibnamefont
  {{Schutz}}},\ }\bibfield  {title} {\bibinfo {title} {{Data analysis of
  gravitational-wave signals from spinning neutron stars: The signal and its
  detection}},\ }\href {https://doi.org/10.1103/PhysRevD.58.063001} {\bibfield
  {journal} {\bibinfo  {journal} {Physical Review D}\ }\textbf {\bibinfo
  {volume} {58}},\ \bibinfo {pages} {063001} (\bibinfo {year} {1998})},\
  \Eprint {https://arxiv.org/abs/gr-qc/9804014} {arXiv:gr-qc/9804014}
  \BibitemShut {NoStop}%
\bibitem [{\citenamefont {{Brady}}\ and\ \citenamefont
  {{Creighton}}(2000)}]{BradCrei2000:SrcPrSrLIIHrrSr:II}%
  \BibitemOpen
  \bibfield  {author} {\bibinfo {author} {\bibfnamefont {P.~R.}\ \bibnamefont
  {{Brady}}}\ and\ \bibinfo {author} {\bibfnamefont {T.}~\bibnamefont
  {{Creighton}}},\ }\bibfield  {title} {\bibinfo {title} {{Searching for
  periodic sources with LIGO. II. Hierarchical searches}},\ }\href
  {https://doi.org/10.1103/PhysRevD.61.082001} {\bibfield  {journal} {\bibinfo
  {journal} {Physical Review D}\ }\textbf {\bibinfo {volume} {61}},\ \bibinfo
  {pages} {082001} (\bibinfo {year} {2000})},\ \Eprint
  {https://arxiv.org/abs/gr-qc/9812014} {arXiv:gr-qc/9812014} \BibitemShut
  {NoStop}%
\bibitem [{\citenamefont {{Wette}}(2015)}]{Wett2015:PrmMASmSrGrvPl}%
  \BibitemOpen
  \bibfield  {author} {\bibinfo {author} {\bibfnamefont {K.}~\bibnamefont
  {{Wette}}},\ }\bibfield  {title} {\bibinfo {title} {{Parameter-space metric
  for all-sky semicoherent searches for gravitational-wave pulsars}},\ }\href
  {https://doi.org/10.1103/PhysRevD.92.082003} {\bibfield  {journal} {\bibinfo
  {journal} {Physical Review D}\ }\textbf {\bibinfo {volume} {92}},\ \bibinfo
  {eid} {082003} (\bibinfo {year} {2015})},\ \Eprint
  {https://arxiv.org/abs/1508.02372} {arXiv:1508.02372 [gr-qc]} \BibitemShut
  {NoStop}%
\bibitem [{\citenamefont {{Wette}}\ \emph
  {et~al.}(2018{\natexlab{a}})\citenamefont {{Wette}}, \citenamefont {{Walsh}},
  \citenamefont {{Prix}},\ and\ \citenamefont
  {{Papa}}}]{WettEtAl2018:ImpSmSCnGrvWUOCnTB}%
  \BibitemOpen
  \bibfield  {author} {\bibinfo {author} {\bibfnamefont {K.}~\bibnamefont
  {{Wette}}}, \bibinfo {author} {\bibfnamefont {S.}~\bibnamefont {{Walsh}}},
  \bibinfo {author} {\bibfnamefont {R.}~\bibnamefont {{Prix}}},\ and\ \bibinfo
  {author} {\bibfnamefont {M.~A.}\ \bibnamefont {{Papa}}},\ }\bibfield  {title}
  {\bibinfo {title} {{Implementing a semicoherent search for continuous
  gravitational waves using optimally constructed template banks}},\ }\href
  {https://doi.org/10.1103/PhysRevD.97.123016} {\bibfield  {journal} {\bibinfo
  {journal} {Physical Review D}\ }\textbf {\bibinfo {volume} {97}},\ \bibinfo
  {pages} {123016} (\bibinfo {year} {2018}{\natexlab{a}})},\ \Eprint
  {https://arxiv.org/abs/1804.03392} {arXiv:1804.03392 [astro-ph.IM]}
  \BibitemShut {NoStop}%
\bibitem [{\citenamefont {{Wette}}\ and\ \citenamefont
  {{Prix}}(2013)}]{WettPrix2013:FPrmMtASrGrvPl}%
  \BibitemOpen
  \bibfield  {author} {\bibinfo {author} {\bibfnamefont {K.}~\bibnamefont
  {{Wette}}}\ and\ \bibinfo {author} {\bibfnamefont {R.}~\bibnamefont
  {{Prix}}},\ }\bibfield  {title} {\bibinfo {title} {{Flat parameter-space
  metric for all-sky searches for gravitational-wave pulsars}},\ }\href
  {https://doi.org/10.1103/PhysRevD.88.123005} {\bibfield  {journal} {\bibinfo
  {journal} {Physical Review D}\ }\textbf {\bibinfo {volume} {88}},\ \bibinfo
  {eid} {123005} (\bibinfo {year} {2013})},\ \Eprint
  {https://arxiv.org/abs/1310.5587} {arXiv:1310.5587 [gr-qc]} \BibitemShut
  {NoStop}%
\bibitem [{\citenamefont {{Wette}}(2016)}]{Wett2016:EmExRVlPrmMASGrvP}%
  \BibitemOpen
  \bibfield  {author} {\bibinfo {author} {\bibfnamefont {K.}~\bibnamefont
  {{Wette}}},\ }\bibfield  {title} {\bibinfo {title} {{Empirically extending
  the range of validity of parameter-space metrics for all-sky searches for
  gravitational-wave pulsars}},\ }\href
  {https://doi.org/10.1103/PhysRevD.94.122002} {\bibfield  {journal} {\bibinfo
  {journal} {Physical Review D}\ }\textbf {\bibinfo {volume} {94}},\ \bibinfo
  {eid} {122002} (\bibinfo {year} {2016})},\ \Eprint
  {https://arxiv.org/abs/1607.00241} {arXiv:1607.00241 [gr-qc]} \BibitemShut
  {NoStop}%
\bibitem [{\citenamefont {{Prix}}(2007)}]{Prix2007:TmpSrGrvWEfLCFPrS}%
  \BibitemOpen
  \bibfield  {author} {\bibinfo {author} {\bibfnamefont {R.}~\bibnamefont
  {{Prix}}},\ }\bibfield  {title} {\bibinfo {title} {{Template-based searches
  for gravitational waves: efficient lattice covering of flat parameter
  spaces}},\ }\href {https://doi.org/10.1088/0264-9381/24/19/S11} {\bibfield
  {journal} {\bibinfo  {journal} {Classical and Quantum Gravity}\ }\textbf
  {\bibinfo {volume} {24}},\ \bibinfo {pages} {S481} (\bibinfo {year}
  {2007})},\ \Eprint {https://arxiv.org/abs/0707.0428} {arXiv:0707.0428
  [gr-qc]} \BibitemShut {NoStop}%
\bibitem [{\citenamefont {{Wette}}(2014)}]{Wett2014:LTmPlcChASrGrvP}%
  \BibitemOpen
  \bibfield  {author} {\bibinfo {author} {\bibfnamefont {K.}~\bibnamefont
  {{Wette}}},\ }\bibfield  {title} {\bibinfo {title} {{Lattice template
  placement for coherent all-sky searches for gravitational-wave pulsars}},\
  }\href {https://doi.org/10.1103/PhysRevD.90.122010} {\bibfield  {journal}
  {\bibinfo  {journal} {Physical Review D}\ }\textbf {\bibinfo {volume} {90}},\
  \bibinfo {pages} {122010} (\bibinfo {year} {2014})},\ \Eprint
  {https://arxiv.org/abs/1410.6882} {arXiv:1410.6882 [gr-qc]} \BibitemShut
  {NoStop}%
\bibitem [{\citenamefont {{Prix}}\ and\ \citenamefont
  {{Shaltev}}(2012)}]{PrixShal2012:SCntGrvWOpStMFCmC}%
  \BibitemOpen
  \bibfield  {author} {\bibinfo {author} {\bibfnamefont {R.}~\bibnamefont
  {{Prix}}}\ and\ \bibinfo {author} {\bibfnamefont {M.}~\bibnamefont
  {{Shaltev}}},\ }\bibfield  {title} {\bibinfo {title} {{Search for continuous
  gravitational waves: Optimal StackSlide method at fixed computing cost}},\
  }\href {https://doi.org/10.1103/PhysRevD.85.084010} {\bibfield  {journal}
  {\bibinfo  {journal} {Physical Review D}\ }\textbf {\bibinfo {volume} {85}},\
  \bibinfo {eid} {084010} (\bibinfo {year} {2012})},\ \Eprint
  {https://arxiv.org/abs/1201.4321} {arXiv:1201.4321 [gr-qc]} \BibitemShut
  {NoStop}%
\bibitem [{\citenamefont {{Manchester}}\ \emph {et~al.}(2005)\citenamefont
  {{Manchester}}, \citenamefont {{Hobbs}}, \citenamefont {{Teoh}},\ and\
  \citenamefont {{Hobbs}}}]{MancEtAl2005:AstTlsNtFcPCt}%
  \BibitemOpen
  \bibfield  {author} {\bibinfo {author} {\bibfnamefont {R.~N.}\ \bibnamefont
  {{Manchester}}}, \bibinfo {author} {\bibfnamefont {G.~B.}\ \bibnamefont
  {{Hobbs}}}, \bibinfo {author} {\bibfnamefont {A.}~\bibnamefont {{Teoh}}},\
  and\ \bibinfo {author} {\bibfnamefont {M.}~\bibnamefont {{Hobbs}}},\
  }\bibfield  {title} {\bibinfo {title} {{The Australia Telescope National
  Facility Pulsar Catalogue}},\ }\href {https://doi.org/10.1086/428488}
  {\bibfield  {journal} {\bibinfo  {journal} {Astronomical Journal}\ }\textbf
  {\bibinfo {volume} {129}},\ \bibinfo {pages} {1993} (\bibinfo {year}
  {2005})},\ \Eprint {https://arxiv.org/abs/astro-ph/0412641}
  {arXiv:astro-ph/0412641 [astro-ph]} \BibitemShut {NoStop}%
\bibitem [{\citenamefont {{Wette}}(2012)}]{Wett2012:EsSnWdpSrGrvP}%
  \BibitemOpen
  \bibfield  {author} {\bibinfo {author} {\bibfnamefont {K.}~\bibnamefont
  {{Wette}}},\ }\bibfield  {title} {\bibinfo {title} {{Estimating the
  sensitivity of wide-parameter-space searches for gravitational-wave
  pulsars}},\ }\href {https://doi.org/10.1103/PhysRevD.85.042003} {\bibfield
  {journal} {\bibinfo  {journal} {Physical Review D}\ }\textbf {\bibinfo
  {volume} {85}},\ \bibinfo {eid} {042003} (\bibinfo {year} {2012})},\ \Eprint
  {https://arxiv.org/abs/1111.5650} {arXiv:1111.5650 [gr-qc]} \BibitemShut
  {NoStop}%
\bibitem [{\citenamefont {{Dreissigacker}}\ \emph {et~al.}(2018)\citenamefont
  {{Dreissigacker}}, \citenamefont {{Prix}},\ and\ \citenamefont
  {{Wette}}}]{DreiEtAl2018:FAcSnsEsCntSr}%
  \BibitemOpen
  \bibfield  {author} {\bibinfo {author} {\bibfnamefont {C.}~\bibnamefont
  {{Dreissigacker}}}, \bibinfo {author} {\bibfnamefont {R.}~\bibnamefont
  {{Prix}}},\ and\ \bibinfo {author} {\bibfnamefont {K.}~\bibnamefont
  {{Wette}}},\ }\bibfield  {title} {\bibinfo {title} {{Fast and accurate
  sensitivity estimation for continuous-gravitational-wave searches}},\ }\href
  {https://doi.org/10.1103/PhysRevD.98.084058} {\bibfield  {journal} {\bibinfo
  {journal} {Physical Review D}\ }\textbf {\bibinfo {volume} {98}},\ \bibinfo
  {pages} {084058} (\bibinfo {year} {2018})}\BibitemShut {NoStop}%
\bibitem [{\citenamefont {{Covas}}\ \emph {et~al.}(2018)\citenamefont
  {{Covas}}, \citenamefont {{Effler}}, \citenamefont {{Goetz}}, \citenamefont
  {{Meyers}}, \citenamefont {{Neunzert}}, \citenamefont {{Oliver}},
  \citenamefont {{Pearlstone}}, \citenamefont {{Roma}}, \citenamefont
  {{Schofield}}, \citenamefont {{Adya}} \emph
  {et~al.}}]{CovaEtAl2018:IdnMtNSArDSPrGrvWFTORAL}%
  \BibitemOpen
  \bibfield  {author} {\bibinfo {author} {\bibfnamefont {P.~B.}\ \bibnamefont
  {{Covas}}}, \bibinfo {author} {\bibfnamefont {A.}~\bibnamefont {{Effler}}},
  \bibinfo {author} {\bibfnamefont {E.}~\bibnamefont {{Goetz}}}, \bibinfo
  {author} {\bibfnamefont {P.~M.}\ \bibnamefont {{Meyers}}}, \bibinfo {author}
  {\bibfnamefont {A.}~\bibnamefont {{Neunzert}}}, \bibinfo {author}
  {\bibfnamefont {M.}~\bibnamefont {{Oliver}}}, \bibinfo {author}
  {\bibfnamefont {B.~L.}\ \bibnamefont {{Pearlstone}}}, \bibinfo {author}
  {\bibfnamefont {V.~J.}\ \bibnamefont {{Roma}}}, \bibinfo {author}
  {\bibfnamefont {R.~M.~S.}\ \bibnamefont {{Schofield}}}, \bibinfo {author}
  {\bibfnamefont {V.~B.}\ \bibnamefont {{Adya}}}, \emph {et~al.},\ }\bibfield
  {title} {\bibinfo {title} {{Identification and mitigation of narrow spectral
  artifacts that degrade searches for persistent gravitational waves in the
  first two observing runs of Advanced LIGO}},\ }\href
  {https://doi.org/10.1103/PhysRevD.97.082002} {\bibfield  {journal} {\bibinfo
  {journal} {Physical Review D}\ }\textbf {\bibinfo {volume} {97}},\ \bibinfo
  {eid} {082002} (\bibinfo {year} {2018})},\ \Eprint
  {https://arxiv.org/abs/1801.07204} {arXiv:1801.07204 [astro-ph.IM]}
  \BibitemShut {NoStop}%
\bibitem [{\citenamefont {{Wette}}\ \emph {et~al.}(2008)\citenamefont {{Wette}}
  \emph {et~al.}}]{WettEtAl2008:SrGrvWvCssLI}%
  \BibitemOpen
  \bibfield  {author} {\bibinfo {author} {\bibfnamefont {K.}~\bibnamefont
  {{Wette}}} \emph {et~al.},\ }\bibfield  {title} {\bibinfo {title} {{Searching
  for gravitational waves from Cassiopeia A with LIGO}},\ }\href
  {https://doi.org/10.1088/0264-9381/25/23/235011} {\bibfield  {journal}
  {\bibinfo  {journal} {Classical and Quantum Gravity}\ }\textbf {\bibinfo
  {volume} {25}},\ \bibinfo {pages} {235011} (\bibinfo {year} {2008})},\
  \Eprint {https://arxiv.org/abs/0802.3332} {arXiv:0802.3332 [gr-qc]}
  \BibitemShut {NoStop}%
\bibitem [{\citenamefont {{Sartore}}\ \emph {et~al.}(2010)\citenamefont
  {{Sartore}}, \citenamefont {{Ripamonti}}, \citenamefont {{Treves}},\ and\
  \citenamefont {{Turolla}}}]{SartEtAl2010:GlcNtStISVlDstDH}%
  \BibitemOpen
  \bibfield  {author} {\bibinfo {author} {\bibfnamefont {N.}~\bibnamefont
  {{Sartore}}}, \bibinfo {author} {\bibfnamefont {E.}~\bibnamefont
  {{Ripamonti}}}, \bibinfo {author} {\bibfnamefont {A.}~\bibnamefont
  {{Treves}}},\ and\ \bibinfo {author} {\bibfnamefont {R.}~\bibnamefont
  {{Turolla}}},\ }\bibfield  {title} {\bibinfo {title} {{Galactic neutron
  stars. I. Space and velocity distributions in the disk and in the halo}},\
  }\href {https://doi.org/10.1051/0004-6361/200912222} {\bibfield  {journal}
  {\bibinfo  {journal} {Astronomy and Astrophysics}\ }\textbf {\bibinfo
  {volume} {510}},\ \bibinfo {eid} {A23} (\bibinfo {year} {2010})},\ \Eprint
  {https://arxiv.org/abs/0908.3182} {arXiv:0908.3182 [astro-ph.GA]}
  \BibitemShut {NoStop}%
\bibitem [{\citenamefont {{Bildsten}}(1998)}]{Bild1998:GrvRdtRtAcNtS}%
  \BibitemOpen
  \bibfield  {author} {\bibinfo {author} {\bibfnamefont {L.}~\bibnamefont
  {{Bildsten}}},\ }\bibfield  {title} {\bibinfo {title} {{Gravitational
  Radiation and Rotation of Accreting Neutron Stars}},\ }\href
  {https://doi.org/10.1086/311440} {\bibfield  {journal} {\bibinfo  {journal}
  {Astrophysical Journal}\ }\textbf {\bibinfo {volume} {501}},\ \bibinfo
  {pages} {L89} (\bibinfo {year} {1998})},\ \Eprint
  {https://arxiv.org/abs/astro-ph/9804325} {arXiv:astro-ph/9804325}
  \BibitemShut {NoStop}%
\bibitem [{\citenamefont {Goetz}(2019)}]{O2-SFT-segments}%
  \BibitemOpen
  \bibfield  {author} {\bibinfo {author} {\bibfnamefont {E.}~\bibnamefont
  {Goetz}} (\bibinfo {collaboration} {{LIGO Scientific Collaboration}}),\
  }\href {https://dcc.ligo.org/LIGO-T1900085/public} {\emph {\bibinfo {title}
  {{Segments used for creating standard SFTs in O2 data}}}},\ \bibinfo {type}
  {Tech. Rep.}\ \bibinfo {number} {T1900085-v1}\ (\bibinfo  {institution}
  {LIGO},\ \bibinfo {year} {2019})\BibitemShut {NoStop}%
\bibitem [{\citenamefont {Dunn}\ \emph {et~al.}(2021)\citenamefont {Dunn} \emph
  {et~al.}}]{GPU-Fstat}%
  \BibitemOpen
  \bibfield  {author} {\bibinfo {author} {\bibfnamefont {L.}~\bibnamefont
  {Dunn}} \emph {et~al.},\ }\href@noop {} {}\bibinfo {howpublished} {in
  preparation} (\bibinfo {year} {2021})\BibitemShut {NoStop}%
\bibitem [{\citenamefont {{Dergachev}}\ and\ \citenamefont
  {{Papa}}(2019)}]{DergPapa2019:SnImpSrPrGrvWUOLD}%
  \BibitemOpen
  \bibfield  {author} {\bibinfo {author} {\bibfnamefont {V.}~\bibnamefont
  {{Dergachev}}}\ and\ \bibinfo {author} {\bibfnamefont {M.~A.}\ \bibnamefont
  {{Papa}}},\ }\bibfield  {title} {\bibinfo {title} {{Sensitivity Improvements
  in the Search for Periodic Gravitational Waves Using O1 LIGO Data}},\ }\href
  {https://doi.org/10.1103/PhysRevLett.123.101101} {\bibfield  {journal}
  {\bibinfo  {journal} {Physical Review Letters}\ }\textbf {\bibinfo {volume}
  {123}},\ \bibinfo {eid} {101101} (\bibinfo {year} {2019})},\ \Eprint
  {https://arxiv.org/abs/1902.05530} {arXiv:1902.05530 [gr-qc]} \BibitemShut
  {NoStop}%
\bibitem [{\citenamefont {{Cutler}}\ and\ \citenamefont
  {{Schutz}}(2005)}]{CutlSchu2005:GnrFsMlDtMGrvWP}%
  \BibitemOpen
  \bibfield  {author} {\bibinfo {author} {\bibfnamefont {C.}~\bibnamefont
  {{Cutler}}}\ and\ \bibinfo {author} {\bibfnamefont {B.~F.}\ \bibnamefont
  {{Schutz}}},\ }\bibfield  {title} {\bibinfo {title} {{Generalized
  $\mathcal{F}$-statistic: Multiple detectors and multiple gravitational wave
  pulsars}},\ }\href {https://doi.org/10.1103/PhysRevD.72.063006} {\bibfield
  {journal} {\bibinfo  {journal} {Physical Review D}\ }\textbf {\bibinfo
  {volume} {72}},\ \bibinfo {pages} {063006} (\bibinfo {year} {2005})},\
  \Eprint {https://arxiv.org/abs/gr-qc/0504011} {arXiv:gr-qc/0504011}
  \BibitemShut {NoStop}%
\bibitem [{\citenamefont {{Abbott}}\ \emph {et~al.}(2009)\citenamefont
  {{Abbott}} \emph {et~al.}}]{LIGO2009:EnsSrPrGrvWvLSD}%
  \BibitemOpen
  \bibfield  {author} {\bibinfo {author} {\bibfnamefont {B.}~\bibnamefont
  {{Abbott}}} \emph {et~al.} (\bibinfo {collaboration} {{LIGO Scientific
  Collaboration}}),\ }\bibfield  {title} {\bibinfo {title} {{Einstein@Home
  search for periodic gravitational waves in LIGO S4 data}},\ }\href
  {https://doi.org/10.1103/PhysRevD.79.022001} {\bibfield  {journal} {\bibinfo
  {journal} {Physical Review D}\ }\textbf {\bibinfo {volume} {79}},\ \bibinfo
  {pages} {022001} (\bibinfo {year} {2009})},\ \Eprint
  {https://arxiv.org/abs/0804.1747} {arXiv:0804.1747 [gr-qc]} \BibitemShut
  {NoStop}%
\bibitem [{\citenamefont {{Parzen}}(1962)}]{Parz1962:EstPrbDnFnMd}%
  \BibitemOpen
  \bibfield  {author} {\bibinfo {author} {\bibfnamefont {E.}~\bibnamefont
  {{Parzen}}},\ }\bibfield  {title} {\bibinfo {title} {{On Estimation of a
  Probability Density Function and Mode}},\ }\href
  {https://doi.org/10.1214/aoms/1177704472} {\bibfield  {journal} {\bibinfo
  {journal} {Annals of Mathematical Statistics}\ }\textbf {\bibinfo {volume}
  {33}},\ \bibinfo {pages} {1065} (\bibinfo {year} {1962})}\BibitemShut
  {NoStop}%
\bibitem [{\citenamefont {{Rosenblatt}}(1956)}]{Rose1956:RmSNnpEstDnFn}%
  \BibitemOpen
  \bibfield  {author} {\bibinfo {author} {\bibfnamefont {M.}~\bibnamefont
  {{Rosenblatt}}},\ }\bibfield  {title} {\bibinfo {title} {{Remarks on Some
  Nonparametric Estimates of a Density Function}},\ }\href
  {https://doi.org/10.1214/aoms/1177728190} {\bibfield  {journal} {\bibinfo
  {journal} {Annals of Mathematical Statistics}\ }\textbf {\bibinfo {volume}
  {27}},\ \bibinfo {pages} {832} (\bibinfo {year} {1956})}\BibitemShut
  {NoStop}%
\bibitem [{\citenamefont {{Abadie}}\ \emph {et~al.}(2010)\citenamefont
  {{Abadie}} \emph {et~al.}}]{LIGO2010:FrSrGrvWYngKNtS}%
  \BibitemOpen
  \bibfield  {author} {\bibinfo {author} {\bibfnamefont {J.}~\bibnamefont
  {{Abadie}}} \emph {et~al.} (\bibinfo {collaboration} {{LIGO Scientific
  Collaboration}}),\ }\bibfield  {title} {\bibinfo {title} {{First Search for
  Gravitational Waves from the Youngest Known Neutron Star}},\ }\href
  {https://doi.org/10.1088/0004-637X/722/2/1504} {\bibfield  {journal}
  {\bibinfo  {journal} {Astrophysical Journal}\ }\textbf {\bibinfo {volume}
  {722}},\ \bibinfo {pages} {1504} (\bibinfo {year} {2010})},\ \Eprint
  {https://arxiv.org/abs/1006.2535} {arXiv:1006.2535 [gr-qc]} \BibitemShut
  {NoStop}%
\bibitem [{\citenamefont {{Messenger}}\ \emph {et~al.}(2015)\citenamefont
  {{Messenger}}, \citenamefont {{Bulten}}, \citenamefont {{Crowder}},
  \citenamefont {{Dergachev}}, \citenamefont {{Galloway}}, \citenamefont
  {{Goetz}}, \citenamefont {{Jonker}}, \citenamefont {{Lasky}}, \citenamefont
  {{Meadors}}, \citenamefont {{Melatos}}, \citenamefont {{Premachandra}},
  \citenamefont {{Riles}}, \citenamefont {{Sammut}}, \citenamefont {{Thrane}},
  \citenamefont {{Whelan}},\ and\ \citenamefont
  {{Zhang}}}]{MessEtAl2015:GrvWSXCmpSMPrDtADt}%
  \BibitemOpen
  \bibfield  {author} {\bibinfo {author} {\bibfnamefont {C.}~\bibnamefont
  {{Messenger}}}, \bibinfo {author} {\bibfnamefont {H.~J.}\ \bibnamefont
  {{Bulten}}}, \bibinfo {author} {\bibfnamefont {S.~G.}\ \bibnamefont
  {{Crowder}}}, \bibinfo {author} {\bibfnamefont {V.}~\bibnamefont
  {{Dergachev}}}, \bibinfo {author} {\bibfnamefont {D.~K.}\ \bibnamefont
  {{Galloway}}}, \bibinfo {author} {\bibfnamefont {E.}~\bibnamefont {{Goetz}}},
  \bibinfo {author} {\bibfnamefont {R.~J.~G.}\ \bibnamefont {{Jonker}}},
  \bibinfo {author} {\bibfnamefont {P.~D.}\ \bibnamefont {{Lasky}}}, \bibinfo
  {author} {\bibfnamefont {G.~D.}\ \bibnamefont {{Meadors}}}, \bibinfo {author}
  {\bibfnamefont {A.}~\bibnamefont {{Melatos}}}, \bibinfo {author}
  {\bibfnamefont {S.}~\bibnamefont {{Premachandra}}}, \bibinfo {author}
  {\bibfnamefont {K.}~\bibnamefont {{Riles}}}, \bibinfo {author} {\bibfnamefont
  {L.}~\bibnamefont {{Sammut}}}, \bibinfo {author} {\bibfnamefont {E.~H.}\
  \bibnamefont {{Thrane}}}, \bibinfo {author} {\bibfnamefont {J.~T.}\
  \bibnamefont {{Whelan}}},\ and\ \bibinfo {author} {\bibfnamefont
  {Y.}~\bibnamefont {{Zhang}}},\ }\bibfield  {title} {\bibinfo {title}
  {{Gravitational waves from Scorpius X-1: A comparison of search methods and
  prospects for detection with advanced detectors}},\ }\href
  {https://doi.org/10.1103/PhysRevD.92.023006} {\bibfield  {journal} {\bibinfo
  {journal} {Physical Review D}\ }\textbf {\bibinfo {volume} {92}},\ \bibinfo
  {eid} {023006} (\bibinfo {year} {2015})},\ \Eprint
  {https://arxiv.org/abs/1504.05889} {arXiv:1504.05889 [gr-qc]} \BibitemShut
  {NoStop}%
\bibitem [{\citenamefont {{Walsh}}\ \emph {et~al.}(2016)\citenamefont
  {{Walsh}}, \citenamefont {{Pitkin}}, \citenamefont {{Oliver}}, \citenamefont
  {{D'Antonio}}, \citenamefont {{Dergachev}}, \citenamefont {{Kr{\'o}lak}},
  \citenamefont {{Astone}}, \citenamefont {{Bejger}}, \citenamefont {{Di
  Giovanni}}, \citenamefont {{Dorosh}}, \citenamefont {{Frasca}}, \citenamefont
  {{Leaci}}, \citenamefont {{Mastrogiovanni}}, \citenamefont {{Miller}},
  \citenamefont {{Palomba}}, \citenamefont {{Papa}}, \citenamefont
  {{Piccinni}}, \citenamefont {{Riles}}, \citenamefont {{Sauter}},\ and\
  \citenamefont {{Sintes}}}]{WalsEtAl2016:CmpMtDtGrvWUnNS}%
  \BibitemOpen
  \bibfield  {author} {\bibinfo {author} {\bibfnamefont {S.}~\bibnamefont
  {{Walsh}}}, \bibinfo {author} {\bibfnamefont {M.}~\bibnamefont {{Pitkin}}},
  \bibinfo {author} {\bibfnamefont {M.}~\bibnamefont {{Oliver}}}, \bibinfo
  {author} {\bibfnamefont {S.}~\bibnamefont {{D'Antonio}}}, \bibinfo {author}
  {\bibfnamefont {V.}~\bibnamefont {{Dergachev}}}, \bibinfo {author}
  {\bibfnamefont {A.}~\bibnamefont {{Kr{\'o}lak}}}, \bibinfo {author}
  {\bibfnamefont {P.}~\bibnamefont {{Astone}}}, \bibinfo {author}
  {\bibfnamefont {M.}~\bibnamefont {{Bejger}}}, \bibinfo {author}
  {\bibfnamefont {M.}~\bibnamefont {{Di Giovanni}}}, \bibinfo {author}
  {\bibfnamefont {O.}~\bibnamefont {{Dorosh}}}, \bibinfo {author}
  {\bibfnamefont {S.}~\bibnamefont {{Frasca}}}, \bibinfo {author}
  {\bibfnamefont {P.}~\bibnamefont {{Leaci}}}, \bibinfo {author} {\bibfnamefont
  {S.}~\bibnamefont {{Mastrogiovanni}}}, \bibinfo {author} {\bibfnamefont
  {A.}~\bibnamefont {{Miller}}}, \bibinfo {author} {\bibfnamefont
  {C.}~\bibnamefont {{Palomba}}}, \bibinfo {author} {\bibfnamefont {M.~A.}\
  \bibnamefont {{Papa}}}, \bibinfo {author} {\bibfnamefont {O.~J.}\
  \bibnamefont {{Piccinni}}}, \bibinfo {author} {\bibfnamefont
  {K.}~\bibnamefont {{Riles}}}, \bibinfo {author} {\bibfnamefont
  {O.}~\bibnamefont {{Sauter}}},\ and\ \bibinfo {author} {\bibfnamefont
  {A.~M.}\ \bibnamefont {{Sintes}}},\ }\bibfield  {title} {\bibinfo {title}
  {{Comparison of methods for the detection of gravitational waves from unknown
  neutron stars}},\ }\href {https://doi.org/10.1103/PhysRevD.94.124010}
  {\bibfield  {journal} {\bibinfo  {journal} {Physical Review D}\ }\textbf
  {\bibinfo {volume} {94}},\ \bibinfo {eid} {124010} (\bibinfo {year}
  {2016})},\ \Eprint {https://arxiv.org/abs/1606.00660} {arXiv:1606.00660
  [gr-qc]} \BibitemShut {NoStop}%
\bibitem [{\citenamefont {{Walsh}}\ \emph {et~al.}(2019)\citenamefont
  {{Walsh}}, \citenamefont {{Wette}}, \citenamefont {{Papa}},\ and\
  \citenamefont {{Prix}}}]{WalsEtAl2019:OpCAnMASCnGrvWEns}%
  \BibitemOpen
  \bibfield  {author} {\bibinfo {author} {\bibfnamefont {S.}~\bibnamefont
  {{Walsh}}}, \bibinfo {author} {\bibfnamefont {K.}~\bibnamefont {{Wette}}},
  \bibinfo {author} {\bibfnamefont {M.~A.}\ \bibnamefont {{Papa}}},\ and\
  \bibinfo {author} {\bibfnamefont {R.}~\bibnamefont {{Prix}}},\ }\bibfield
  {title} {\bibinfo {title} {{Optimizing the choice of analysis method for
  all-sky searches for continuous gravitational waves with Einstein@Home}},\
  }\href {https://doi.org/10.1103/PhysRevD.99.082004} {\bibfield  {journal}
  {\bibinfo  {journal} {Physical Review D}\ }\textbf {\bibinfo {volume} {99}},\
  \bibinfo {pages} {082004} (\bibinfo {year} {2019})}\BibitemShut {NoStop}%
\bibitem [{\citenamefont {{Ming}}\ \emph {et~al.}(2016)\citenamefont {{Ming}},
  \citenamefont {{Krishnan}}, \citenamefont {{Papa}}, \citenamefont
  {{Aulbert}},\ and\ \citenamefont {{Fehrmann}}}]{MingEtAl2016:OpDrSrCntGrvW}%
  \BibitemOpen
  \bibfield  {author} {\bibinfo {author} {\bibfnamefont {J.}~\bibnamefont
  {{Ming}}}, \bibinfo {author} {\bibfnamefont {B.}~\bibnamefont {{Krishnan}}},
  \bibinfo {author} {\bibfnamefont {M.~A.}\ \bibnamefont {{Papa}}}, \bibinfo
  {author} {\bibfnamefont {C.}~\bibnamefont {{Aulbert}}},\ and\ \bibinfo
  {author} {\bibfnamefont {H.}~\bibnamefont {{Fehrmann}}},\ }\bibfield  {title}
  {\bibinfo {title} {{Optimal directed searches for continuous gravitational
  waves}},\ }\href {https://doi.org/10.1103/PhysRevD.93.064011} {\bibfield
  {journal} {\bibinfo  {journal} {Physical Review D}\ }\textbf {\bibinfo
  {volume} {93}},\ \bibinfo {eid} {064011} (\bibinfo {year} {2016})},\ \Eprint
  {https://arxiv.org/abs/1510.03417} {arXiv:1510.03417 [gr-qc]} \BibitemShut
  {NoStop}%
\bibitem [{\citenamefont {{Dergachev}}\ and\ \citenamefont
  {{Papa}}(2020)}]{DergPapa2020:RsFASCnGrvWSmlSr}%
  \BibitemOpen
  \bibfield  {author} {\bibinfo {author} {\bibfnamefont {V.}~\bibnamefont
  {{Dergachev}}}\ and\ \bibinfo {author} {\bibfnamefont {M.~A.}\ \bibnamefont
  {{Papa}}},\ }\bibfield  {title} {\bibinfo {title} {{Results from the First
  All-Sky Search for Continuous Gravitational Waves from Small-Ellipticity
  Sources}},\ }\href {https://doi.org/10.1103/PhysRevLett.125.171101}
  {\bibfield  {journal} {\bibinfo  {journal} {Physical Review Letters}\
  }\textbf {\bibinfo {volume} {125}},\ \bibinfo {eid} {171101} (\bibinfo {year}
  {2020})},\ \Eprint {https://arxiv.org/abs/2004.08334} {arXiv:2004.08334
  [gr-qc]} \BibitemShut {NoStop}%
\bibitem [{\citenamefont {Dergachev}\ and\ \citenamefont
  {Papa}(2020)}]{dergachev2020results}%
  \BibitemOpen
  \bibfield  {author} {\bibinfo {author} {\bibfnamefont {V.}~\bibnamefont
  {Dergachev}}\ and\ \bibinfo {author} {\bibfnamefont {M.~A.}\ \bibnamefont
  {Papa}},\ }\href@noop {} {\bibinfo {title} {{Results from high-frequency
  all-sky search for continuous gravitational waves from small-ellipticity
  sources}}} (\bibinfo {year} {2020}),\ \Eprint
  {https://arxiv.org/abs/2012.04232} {arXiv:2012.04232 [gr-qc]} \BibitemShut
  {NoStop}%
\bibitem [{\citenamefont {{Bonazzola}}\ and\ \citenamefont
  {{Gourgoulhon}}(1996)}]{BonaGour1996:GrvWPlEmMgFIDst}%
  \BibitemOpen
  \bibfield  {author} {\bibinfo {author} {\bibfnamefont {S.}~\bibnamefont
  {{Bonazzola}}}\ and\ \bibinfo {author} {\bibfnamefont {E.}~\bibnamefont
  {{Gourgoulhon}}},\ }\bibfield  {title} {\bibinfo {title} {{Gravitational
  waves from pulsars: emission by the magnetic field induced distortion}},\
  }\href {https://arxiv.org/abs/astro-ph/9602107} {\bibfield  {journal}
  {\bibinfo  {journal} {Astronomy \& Astrophysics}\ }\textbf {\bibinfo {volume}
  {312}},\ \bibinfo {pages} {675} (\bibinfo {year} {1996})},\ \Eprint
  {https://arxiv.org/abs/astro-ph/9602107} {arXiv:astro-ph/9602107}
  \BibitemShut {NoStop}%
\bibitem [{\citenamefont {{LIGO Scientific Collaboration}}(2018)}]{lalsuite}%
  \BibitemOpen
  \bibfield  {author} {\bibinfo {author} {\bibnamefont {{LIGO Scientific
  Collaboration}}},\ }\href {https://doi.org/10.7935/GT1W-FZ16} {\bibinfo
  {title} {{LIGO} {A}lgorithm {L}ibrary - {LALS}uite}},\ \bibinfo
  {howpublished} {Free software (GPL)} (\bibinfo {year} {2018})\BibitemShut
  {NoStop}%
\bibitem [{\citenamefont {Eaton}\ \emph {et~al.}(2020)\citenamefont {Eaton},
  \citenamefont {Bateman}, \citenamefont {Hauberg},\ and\ \citenamefont
  {Wehbring}}]{octave}%
  \BibitemOpen
  \bibfield  {author} {\bibinfo {author} {\bibfnamefont {J.~W.}\ \bibnamefont
  {Eaton}}, \bibinfo {author} {\bibfnamefont {D.}~\bibnamefont {Bateman}},
  \bibinfo {author} {\bibfnamefont {S.}~\bibnamefont {Hauberg}},\ and\ \bibinfo
  {author} {\bibfnamefont {R.}~\bibnamefont {Wehbring}},\ }\href
  {https://www.gnu.org/software/octave/doc/} {\emph {\bibinfo {title} {{GNU
  Octave manual: a high-level interactive language for numerical
  computations}}}} (\bibinfo {year} {2020})\BibitemShut {NoStop}%
\bibitem [{\citenamefont {{Wette}}\ \emph
  {et~al.}(2018{\natexlab{b}})\citenamefont {{Wette}}, \citenamefont {{Prix}},
  \citenamefont {{Keitel}}, \citenamefont {{Pitkin}}, \citenamefont
  {{Dreissigacker}}, \citenamefont {{Whelan}},\ and\ \citenamefont
  {{Leaci}}}]{WettEtAl2018:OcLOFnCntGrvDAn}%
  \BibitemOpen
  \bibfield  {author} {\bibinfo {author} {\bibfnamefont {K.}~\bibnamefont
  {{Wette}}}, \bibinfo {author} {\bibfnamefont {R.}~\bibnamefont {{Prix}}},
  \bibinfo {author} {\bibfnamefont {D.}~\bibnamefont {{Keitel}}}, \bibinfo
  {author} {\bibfnamefont {M.}~\bibnamefont {{Pitkin}}}, \bibinfo {author}
  {\bibfnamefont {C.}~\bibnamefont {{Dreissigacker}}}, \bibinfo {author}
  {\bibfnamefont {J.~T.}\ \bibnamefont {{Whelan}}},\ and\ \bibinfo {author}
  {\bibfnamefont {P.}~\bibnamefont {{Leaci}}},\ }\bibfield  {title} {\bibinfo
  {title} {{OctApps: a library of Octave functions for continuous
  gravitational-wave data analysis}},\ }\href
  {https://doi.org/10.21105/joss.00707} {\bibfield  {journal} {\bibinfo
  {journal} {Journal of Open Source Software}\ }\textbf {\bibinfo {volume}
  {3}},\ \bibinfo {pages} {707} (\bibinfo {year}
  {2018}{\natexlab{b}})}\BibitemShut {NoStop}%
\bibitem [{\citenamefont {{Python Software Foundation}}(2020)}]{python}%
  \BibitemOpen
  \bibfield  {author} {\bibinfo {author} {\bibnamefont {{Python Software
  Foundation}}},\ }\href {https://docs.python.org/reference/} {\emph {\bibinfo
  {title} {{Python Language Reference}}}} (\bibinfo {year} {2020})\BibitemShut
  {NoStop}%
\bibitem [{\citenamefont {Harris}\ \emph {et~al.}(2020)\citenamefont {Harris},
  \citenamefont {Millman}, \citenamefont {van~der Walt}, \citenamefont
  {Gommers}, \citenamefont {Virtanen}, \citenamefont {Cournapeau},
  \citenamefont {Wieser}, \citenamefont {Taylor}, \citenamefont {Berg},
  \citenamefont {Smith} \emph {et~al.}}]{numpy}%
  \BibitemOpen
  \bibfield  {author} {\bibinfo {author} {\bibfnamefont {C.~R.}\ \bibnamefont
  {Harris}}, \bibinfo {author} {\bibfnamefont {K.~J.}\ \bibnamefont {Millman}},
  \bibinfo {author} {\bibfnamefont {S.~J.}\ \bibnamefont {van~der Walt}},
  \bibinfo {author} {\bibfnamefont {R.}~\bibnamefont {Gommers}}, \bibinfo
  {author} {\bibfnamefont {P.}~\bibnamefont {Virtanen}}, \bibinfo {author}
  {\bibfnamefont {D.}~\bibnamefont {Cournapeau}}, \bibinfo {author}
  {\bibfnamefont {E.}~\bibnamefont {Wieser}}, \bibinfo {author} {\bibfnamefont
  {J.}~\bibnamefont {Taylor}}, \bibinfo {author} {\bibfnamefont
  {S.}~\bibnamefont {Berg}}, \bibinfo {author} {\bibfnamefont {N.~J.}\
  \bibnamefont {Smith}}, \emph {et~al.},\ }\bibfield  {title} {\bibinfo {title}
  {{Array programming with {NumPy}}},\ }\href
  {https://doi.org/10.1038/s41586-020-2649-2} {\bibfield  {journal} {\bibinfo
  {journal} {Nature}\ }\textbf {\bibinfo {volume} {585}},\ \bibinfo {pages}
  {357–362} (\bibinfo {year} {2020})}\BibitemShut {NoStop}%
\bibitem [{\citenamefont {Hunter}(2007)}]{matplotlib}%
  \BibitemOpen
  \bibfield  {author} {\bibinfo {author} {\bibfnamefont {J.~D.}\ \bibnamefont
  {Hunter}},\ }\bibfield  {title} {\bibinfo {title} {{Matplotlib: A 2D graphics
  environment}},\ }\href {https://doi.org/10.1109/MCSE.2007.55} {\bibfield
  {journal} {\bibinfo  {journal} {Computing in Science \& Engineering}\
  }\textbf {\bibinfo {volume} {9}},\ \bibinfo {pages} {90} (\bibinfo {year}
  {2007})}\BibitemShut {NoStop}%
\bibitem [{\citenamefont {Virtanen}\ \emph {et~al.}(2020)\citenamefont
  {Virtanen}, \citenamefont {Gommers}, \citenamefont {Oliphant}, \citenamefont
  {Haberland}, \citenamefont {Reddy}, \citenamefont {Cournapeau}, \citenamefont
  {Burovski}, \citenamefont {Peterson}, \citenamefont {Weckesser},
  \citenamefont {Bright} \emph {et~al.}}]{scipy}%
  \BibitemOpen
  \bibfield  {author} {\bibinfo {author} {\bibfnamefont {P.}~\bibnamefont
  {Virtanen}}, \bibinfo {author} {\bibfnamefont {R.}~\bibnamefont {Gommers}},
  \bibinfo {author} {\bibfnamefont {T.~E.}\ \bibnamefont {Oliphant}}, \bibinfo
  {author} {\bibfnamefont {M.}~\bibnamefont {Haberland}}, \bibinfo {author}
  {\bibfnamefont {T.}~\bibnamefont {Reddy}}, \bibinfo {author} {\bibfnamefont
  {D.}~\bibnamefont {Cournapeau}}, \bibinfo {author} {\bibfnamefont
  {E.}~\bibnamefont {Burovski}}, \bibinfo {author} {\bibfnamefont
  {P.}~\bibnamefont {Peterson}}, \bibinfo {author} {\bibfnamefont
  {W.}~\bibnamefont {Weckesser}}, \bibinfo {author} {\bibfnamefont
  {J.}~\bibnamefont {Bright}}, \emph {et~al.},\ }\bibfield  {title} {\bibinfo
  {title} {{{SciPy} 1.0: Fundamental Algorithms for Scientific Computing in
  Python}},\ }\href {https://doi.org/10.1038/s41592-019-0686-2} {\bibfield
  {journal} {\bibinfo  {journal} {Nature Methods}\ }\textbf {\bibinfo {volume}
  {17}},\ \bibinfo {pages} {261} (\bibinfo {year} {2020})}\BibitemShut
  {NoStop}%
\end{thebibliography}
\end{document}